\documentclass[english,aps,prb,superscriptaddress,reprint,longbibliograhy]{revtex4-2}
\usepackage[T1]{fontenc}
\usepackage[latin9]{inputenc}
\usepackage{babel}
\usepackage{graphicx}
\usepackage{bm}
\usepackage{amsmath}
\usepackage{amssymb}
\usepackage{amsfonts}
\usepackage{dcolumn}
\usepackage{bbold}
\usepackage{tikz}
\usepackage{xfp}
\usepackage{verbatim}

\usepackage[margin=15mm]{geometry}
\colorlet{linkequation}{blue}
\usepackage[colorlinks=true,linkcolor=blue,citecolor=red,urlcolor=blue]{hyperref}
\usetikzlibrary{shadings}

\usepackage{color}

\begin{document}

\begin{abstract}

Inertial effects in spin dynamics are theoretically predicted to emerge at ultrashort time scales, but their experimental signatures are often ambiguous. Here, we calculate the spin-wave spectrum in ferromagnets and two-sublattice antiferromagnets in the presence of inertial effects. It is shown how precession and nutation spin waves hybridize with each other, leading to the renormalization of the frequencies, the group velocities, the effective gyromagnetic ratios and the effective damping parameters. Possible ways of distinguishing between the signatures of inertial dynamics and similar effects explainable within conventional models are discussed.    

\end{abstract}

\title{Inertial spin waves in ferromagnets and antiferromagnets}

\author{Ritwik Mondal}
\email[]{ritwik@iitism.ac.in}
\affiliation{Department of Physics, Indian Institute of Technology (ISM) Dhanbad, IN-826004, Dhanbad, India}

\author{Levente R\'ozsa}
\email[]{levente.rozsa@uni-konstanz.de}
\affiliation{Fachbereich Physik, Universit\"at Konstanz, DE-78457 Konstanz, Germany}

\maketitle
\date{\today}

\section{Introduction}

Spin waves or magnons correspond to the elementary excitations of magnetically ordered systems. The magnon dispersion relation provides extensive information about the spin configuration and about the microscopic magnetic interactions. The dispersion relation is possible to probe using a variety of experimental methods including light, spin-polarized electron or neutron scattering. Magnons have also been put forward as possible information carriers in computational architectures~\cite{Klinger2015,Abdulqader2020}, where their typically short wavelengths at the required operational frequencies facilitate their integration with existing CMOS devices.

Combining a high speed with a long lifetime of magnons is preferred for such computing applications, but these requirements are often competing with each other. Very low damping parameters and consequently long lifetimes have been achieved in ferromagnets or strongly uncompensated ferrimagnets such as yttrium iron garnet, but the typical excitation frequencies of these materials lie in the GHz regime~\cite{Kittel1948,Podler1949,Farle1998,Tserkovnyak2005,John2017}. Antiferromagnetic materials are characterized by THz excitation frequencies due to the exchange enhancement caused by the coupling between the sublattices. However, the exchange enhancement also affects the effective damping parameters~\cite{Gurevich}, leading to short magnon lifetimes.

A possible solution for achieving high frequencies and low damping is offered by inertial spin dynamics~\cite{Suhl1998,Ciornei2011,Henk2012}. Including an inertial term in the Landau--Lifshitz--Gilbert equation~\cite{landau35,gilbert04,nowak2007handbook} describing the time evolution of the spins leads to a separation between the directions of the magnetic moment and of the angular momentum, giving rise to spin nutation. Various theoretical proposals have been put forward for explaining the microscopic origin of the inertial term~\cite{Ciornei2011,Bhattacharjee2012,Fahnle2011,fahnle2013erratum,Mondal2017Nutation,Mondal2018JPCM,Giordano2020,Titov2021}. Theoretical and experimental estimates for the inertial relaxation time $\eta$ (see Eq.~\eqref{Eq2} below for the definition used here) range from a few fs to several hundred fs~\cite{Ciornei2011,Bhattacharjee2012,Henk2012,Li2015,Thonig2017,neeraj2019experimental}, establishing the characteristic time scale on which nutation can be observed.

The inertial dynamics introduces an additional peak in the ferromagnetic resonance (FMR) spectrum. While the traditional precession resonance peak is observable at GHz frequencies, the nutation resonance occurs in the THz range~\cite{Olive2012,Olive2015,MondalJPCM2021} based on the above estimates for the inertial relaxation time. The nutation resonance has been observed experimentally in bulk CoFeB and NiFe \cite{neeraj2019experimental} and in epitaxial Co films \cite{unikandanunni2021inertial}. Despite the enhancement of the excitation frequency, the effective damping parameter characterizing the linewidth of the resonance is reduced due to the inertial dynamics~\cite{Olive2015,Mondal2020nutation}. The presence of spin inertia not only introduces a second resonance peak, but also shifts the precession resonance frequency to a lower value~\cite{cherkasskii2020nutation,Mondal2021PRB}. While the spins rotate counterclockwise around their equilibrium direction at the precession resonance, they rotate clockwise at the nutation resonance~\cite{Kikuchi,Mondal2021PRBSpinCurrent}. The opposite rotational sense also induces a sign change in the injected spin current from a ferromagnet to an adjacent metallic layer between the precession and the nutation resonances~\cite{Mondal2021PRBSpinCurrent}. Compared to ferromagnets, the interplay between precession and nutation is exchange enhanced in two-sublattice antiferromagnets, causing an increase in the frequency and magnitude of the nutation resonance peak while the precession resonance is suppressed~\cite{Mondal2020nutation}. The high susceptibility in FMR of the nutation resonance compared to the precession resonance in antiferromagnets is explained by the profile of the excitation modes: while the two sublattices align almost antiparallel at the precession resonance, the nutation resonance is dominated by one of the sublattices, thereby a finite net magnetic moment arises~\cite{Mondal2021PRBSpinCurrent}.

Away from the center of the Brillouin zone probed by FMR, the inertial dynamics gives rise to propagating nutation spin waves at finite wave vectors, which could be detected, e.g., by Brillouin Light Scattering or spin-polarized electron or neutron scattering. It has been derived in previous theoretical works~\cite{Makhfudz2020,Cherkasskii2021,PhysRevB.105.214414,PhysRevB.104.054425} that excitation frequencies of nutation spin waves are enhanced compared to conventional, or precession, spin waves in ferromagnets. These studies either primarily focused on small wave vectors where magnetostatic effects dominate~\cite{Cherkasskii2021,PhysRevB.105.214414}, or only took isotropic exchange interactions into account~\cite{Makhfudz2020,PhysRevB.104.054425}. Nutation spin waves in antiferromagnets have not been considered so far.

Going beyond the isotropic exchange, the Dzyaloshinsky--Moriya interaction~\cite{Dzyaloshinsky,Moriya} arises in systems with broken inversion symmetry due to spin--orbit coupling, and leads to non-reciprocal magnon propagation in ferromagnets, i.e., a different frequency of magnons with wave vectors $\boldsymbol{k}$ and $-\boldsymbol{k}$~\cite{Coldea,Udvardi,Costa}. This asymmetry in the spectrum is widely used for the experimental determination of the Dzyaloshinsky--Moriya interaction~\cite{Zakeri,Kuepferling2022}. This interaction is essential for understanding the stabilization of non-collinear spin structures such as chiral domain walls, spin spirals and skyrmions in systems both with ferromagnetic and antiferromagnetic isotropic exchange interactions~\cite{Bergmann2014}, as well as for the formation of topological magnon boundary states~\cite{Zhang2013}. However, the effect of the Dzyaloshinsky--Moriya interaction on nutation spin waves remains unexplored.

Here, we present a calculation of the magnon frequencies in the presence of inertial effects within linear spin-wave theory, including the effects of Heisenberg exchange, Dzyaloshinsky--Moriya interaction and magnetocrystalline anisotropy terms in the Hamiltonian. The results are discussed for ferromagnets in Sec.~\ref{sec2} and for antiferromagnets in Sec.~\ref{sec3}. The most pronounced signature of the inertial dynamics is the emergence of high-frequency nutation magnon bands. In order to account for the difficulties in the experimental detection of spin waves at such high frequencies, it is also discussed how the signatures of inertial dynamics could be observed in the group velocity, the gyromagnetic ratio or the effective damping parameter of precession magnons due to their hybridization with nutation spin waves.

\section{Ferromagnets\label{sec2}}
\subsection{Linear spin-wave theory}
We will consider the classical atomistic spin Hamiltonian
\begin{align}
    \mathcal{H} & = \frac{1}{2}\sum_{i\neq j} \boldsymbol{S}_i \boldsymbol{J}_{ij} \boldsymbol{S}_j + \sum_{i} \boldsymbol{S}_i \boldsymbol{K}_i\boldsymbol{S}_i - \sum_i \boldsymbol{B}_i M_i\boldsymbol{S}_i,
    \label{Eq4}
\end{align}
where $\boldsymbol{S}_i$ is the unit magnetization vector and $M_{i}$ is the magnitude of the moment at lattice site $i$, the $3\times 3$ tensors $\boldsymbol{J}_{ij}$ and $\boldsymbol{K}_i$ describe interactions between the spins and on-site magnetocrystalline anisotropy, respectively, while $\boldsymbol{B}_i$ stands for the external magnetic field.

The inertial Landau--Lifshitz--Gilbert (ILLG) equation describing the time evolution of the spins reads \cite{Ciornei2011,Olive2012,Mondal2017Nutation}
\begin{align}
    \frac{\textrm{d} \boldsymbol{S}_i}{\textrm{d} t} & = \boldsymbol{S}_i \times \left[ -\gamma_i  \boldsymbol{B}^{\rm eff}_i + \alpha_i \frac{\textrm{d} \boldsymbol{S}_i}{\textrm{d} t} + \eta_i \frac{\textrm{d}^2 \boldsymbol{S}_i}{\textrm{d} t^2}\right],
    \label{Eq2}
\end{align}
where $\gamma_{i}$ is the absolute value of the gyromagnetic ratio for electrons, $\alpha_{i}$ is the Gilbert damping, $\eta_{i}$ is the inertial relaxation time, and $\boldsymbol{B}^{\rm eff}_i=-\frac{1}{M_{i}}\frac{\partial \mathcal{H}}{\partial \boldsymbol{S}_i}$ is the effective magnetic field acting on the spins and the effective Hamiltonian of the total system is denoted as $\mathcal{H}$.

In the ferromagnetic ground state, we assume that all spins are pointing along the $\hat{\bm{z}}$ direction. We introduce the variables $\beta_{1i}$ and $\beta_{2i}$ to describe small deviations from the ground state. The spin directions are expanded up to second order in these variables as \cite{Rozsa_2013}
\begin{align}
    \boldsymbol{S}_i & = \begin{pmatrix}
    {\beta}_{2i}\\
    - {\beta}_{1i}\\
    1 - \dfrac{{\beta}_{1i}^2}{2} - \dfrac{{\beta}_{2i}^2}{2}
    \end{pmatrix}.
    \label{Eq3}
\end{align}
This expansion enables the linearization of the ILLG Eq.~\eqref{Eq2}, see Appendix~\ref{AppendixA} for details. In the following, we consider translationally invariant systems with a single sublattice, meaning $\gamma_{i}=\gamma$, $\alpha_{i}=\alpha$, $\eta_{i}=\eta$, $M_{i}=M$, $\boldsymbol{K}_{i}=\boldsymbol{K}$, $\boldsymbol{J}_{ij}=\boldsymbol{J}\left(\boldsymbol{R}_{i}-\boldsymbol{R}_{j}\right)$ and homogeneous external magnetic field $\boldsymbol{B}_{i}=\boldsymbol{B}$. We introduce the spatial Fourier transforms of the small variables $\tilde{\beta}_{1,2}\left(\boldsymbol{k}\right)$, and use the circularly polarized basis defined by $\tilde{\beta}_{\pm}\left(\boldsymbol{k}\right) = \tilde{\beta}_{2}\left(\boldsymbol{k}\right) \pm {\rm i}  \tilde{\beta}_{1}\left(\boldsymbol{k}\right)$. We also assume a harmonic time dependence for the variables $\tilde{\beta}_{\pm}\left(\boldsymbol{k}\right)$, replacing the differential operator $-\textrm{i}\frac{\textrm{d}}{\textrm{d}t}$ with the frequency $\omega$, turning the ILLG dynamical equations into a set of algebraic equations. These may be written as
\begin{align}
&\left[D\left(\omega\right)+H_{\textrm{SW}}\left(\boldsymbol{k}\right)\right]\tilde{\boldsymbol{\beta}}\left(\boldsymbol{k}\right)=0,
\label{Eq10}
\end{align}
with
\begin{align}
    \tilde{\boldsymbol{\beta}}\left(\boldsymbol{k}\right)=&\begin{pmatrix}\tilde{\beta}_{+}\left(\boldsymbol{k}\right) & \tilde{\beta}_{-}\left(\boldsymbol{k}\right)\end{pmatrix}^{T},\label{Eq10b}\\
    D\left(\omega\right)=&-\eta\omega^{2}+\left(\textrm{i}\alpha+\sigma^{z}\right)\omega,\label{Eq10c}
\end{align}
where $\sigma^{z}$ is the Pauli matrix acting on the components of $\tilde{\boldsymbol{\beta}}\left(\boldsymbol{k}\right)$ and the spin-wave Hamiltonian $H_{\textrm{SW}}\left(\boldsymbol{k}\right)$ is given in Appendix~\ref{AppendixA}. The solutions of the linearized ILLG equation form pairs: for each eigenvalue $\omega\left(\boldsymbol{k}\right)$ and eigenvector $\tilde{\boldsymbol{\beta}}\left(\boldsymbol{k}\right)$, there is a corresponding eigenvalue $\omega\left(-\boldsymbol{k}\right)=-\omega^{*}\left(\boldsymbol{k}\right)$ and eigenvector $\tilde{\boldsymbol{\beta}}\left(-\boldsymbol{k}\right)=\mathcal{C}\tilde{\boldsymbol{\beta}}\left(\boldsymbol{k}\right)$, where $\mathcal{C}=\sigma^{x}\mathcal{K}$ is the particle-hole symmetry operator, $\sigma^{x}$ is a Pauli matrix and $\mathcal{K}$ is complex conjugation. Particle-hole symmetry describes a special structure of the dynamical equation~\cite{Flynn2020}, which is enforced by the commutation relations of bosonic creation and annihilation operators in the quantum-mechanical case, and by the requirement that the spin-wave eigenfunctions $\beta_{1i},\beta_{2i}$ can always be chosen to be real-valued in the classical model discussed here. Here it is demonstrated that this symmetry is conserved even in the presence of inertial and damping terms which do not have a direct quantum-mechanical analogue.
\begin{figure}
    \centering
    \includegraphics[scale = 0.44]{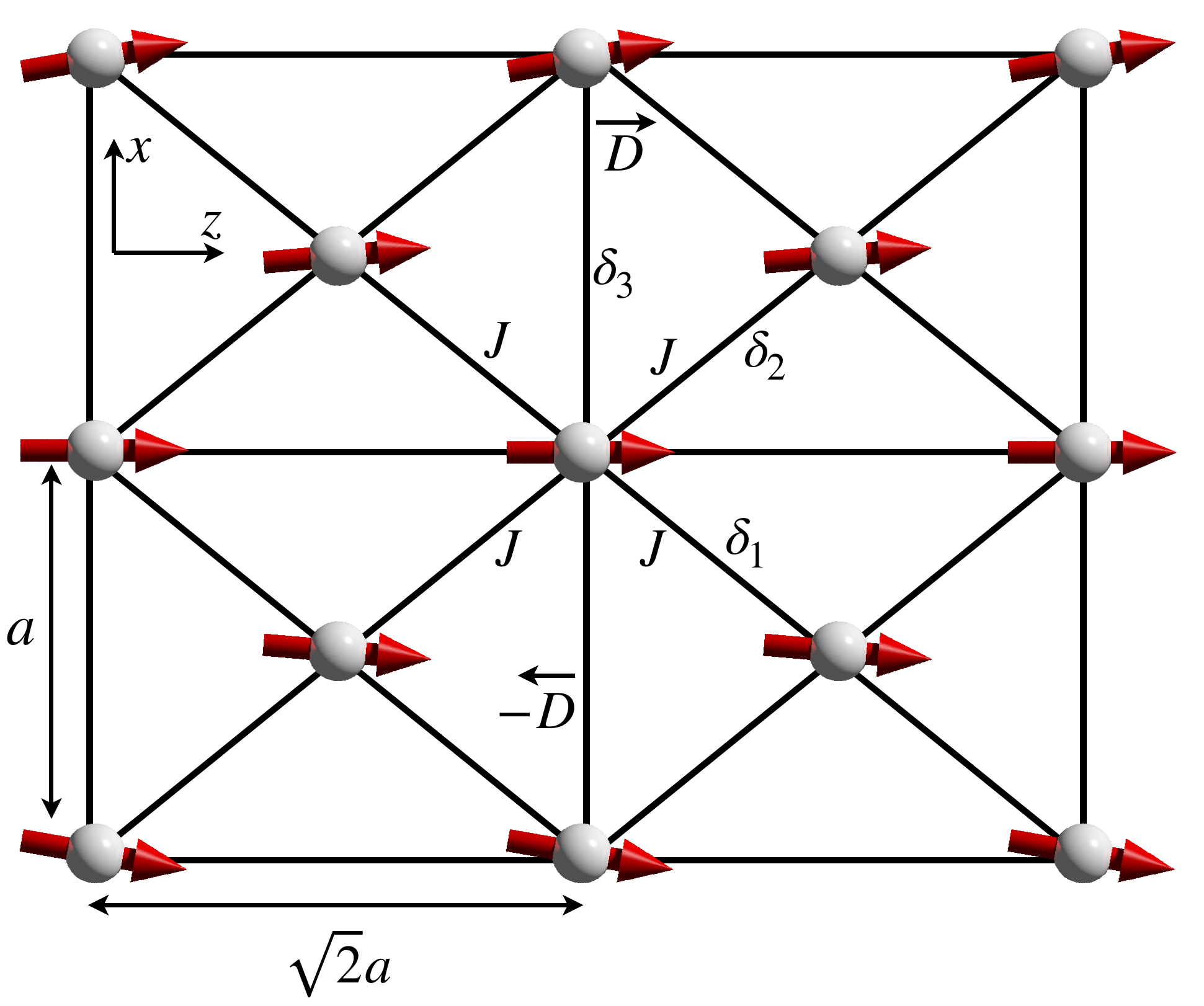}
    \caption{Sketch of a two-dimensional centered rectangular lattice, describing a magnetic monolayer on a bcc(110) surface. The lattice constants are $\sqrt{2}a$ and $a$ along the $z$ and $x$ directions, respectively. The arrows show the ground state spin orientation along the equilibrium direction $z$, with a small tilting of spins illustrating the traveling spin wave. The nearest-neighbor lattice vectors are labeled as $\boldsymbol{\delta}_1$ and $\boldsymbol{\delta}_2$, while the next-nearest-neighbor lattice vector is $\boldsymbol{\delta}_3$. The nearest-neighbor coupling coefficients are denoted by $J$ containing $J^{xx}=J^{yy}$ and $J^{zz}$, while the next-nearest neighbor Dzyaloshinsky-Moriya vectors are $\boldsymbol{D}$. }
    \label{Fig1}
\end{figure}
\begin{figure*}[tbh!]
    \centering
    \includegraphics[scale = 0.25]{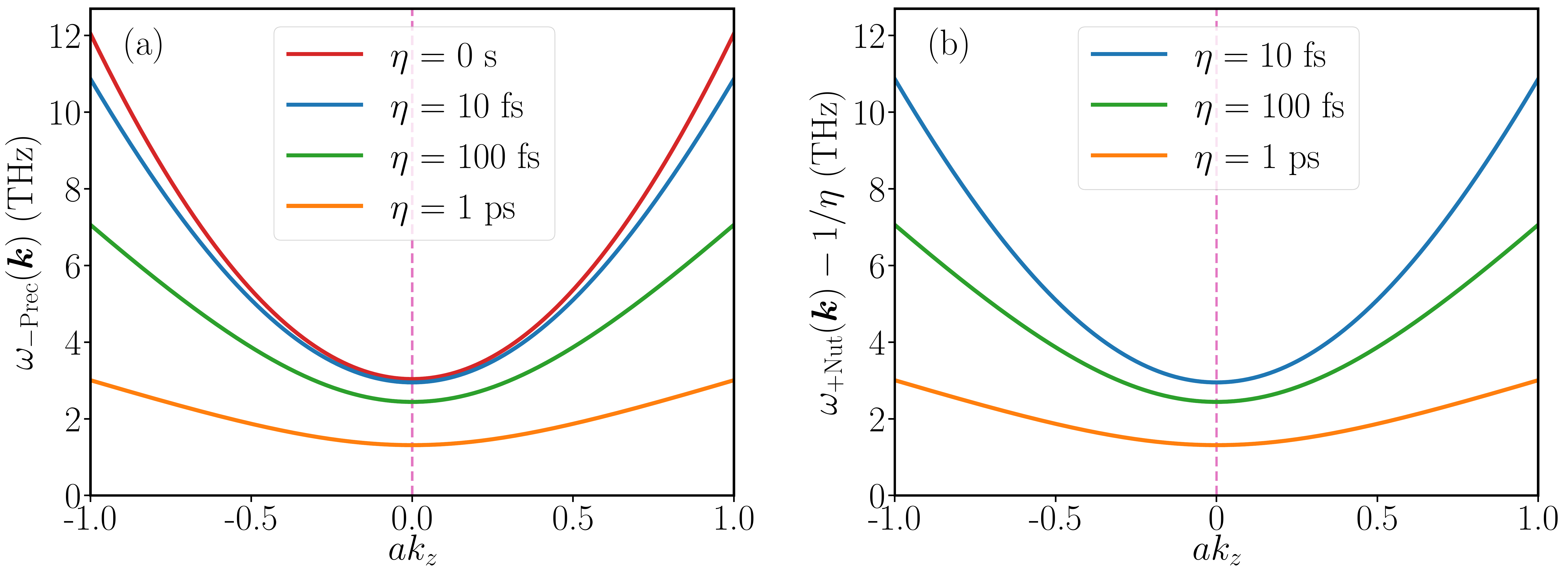}
    \vfill
    \includegraphics[scale = 0.25]{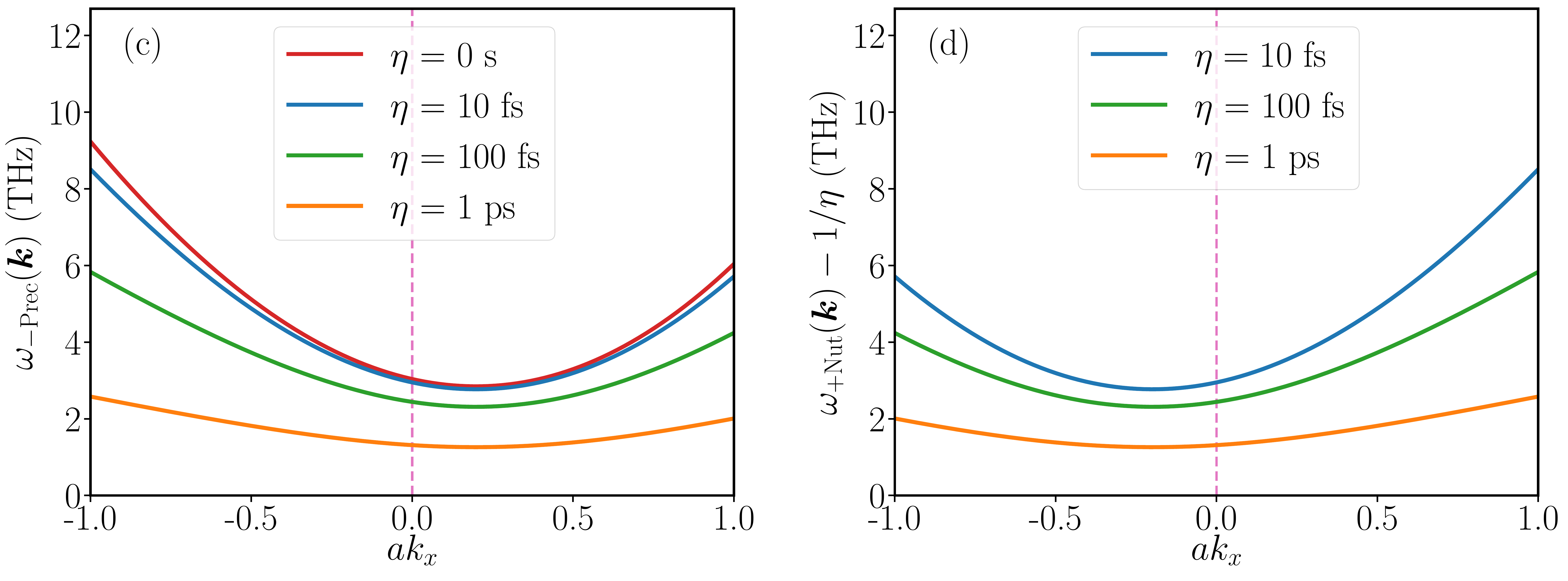}
    \caption{Spin-wave dispersion in a ferromagnet without and with the inertia. (a) Precession and  (b) nutation bands along $k_z$ for $k_x = 0$. (c) Precession and  (d) nutation bands along $k_x$ for $k_z = 0$. The parameters are  $J^{zz} = -1.02\times 10^{-21}$ J, $J^{xx} = -0.99\times 10^{-21}$ J, $D = 10^{-22}$ J, $K^{zz} = -10^{-22}$ J, $B^{z} = 0$ and $\alpha = 0$.}
    \label{Fig2}
\end{figure*}

\subsection{Dispersion relation}
As a specific example, we consider the system sketched in Fig.~\ref{Fig1}, representing a magnetic monolayer on a bcc(110) surface, for example Fe/W(110) where the prominent effect of the Dzyaloshinsky--Moriya interaction on the spin-wave spectrum has been established in previous works~\cite{Zakeri,Udvardi,Rozsa_2013}. The $z$ direction describing the orientation of the spins is oriented along the long side of the centered rectangular unit cell, while $x$ is along the perpendicular in-plane direction.  
We will consider on-site anisotropy with only $K^{zz}$ being finite, nearest-neighbor exchange interactions with $J^{zz}$ and $J^{xx}=J^{yy}$, and next-nearest-neighbor Dzyaloshinsky--Moriya interactions with $D_{ij}=J_{ij}^{xy}=-J_{ij}^{yx}$. The symmetry of the system ensures that the Dzyaloshinsky--Moriya vector $\boldsymbol{D}$ is oriented along the $z$ direction for next-nearest neighbors, and only this component of the vector enters the linearized ILLG equation. Under these assumptions, Eq.~\eqref{Eq10} simplifies to two uncoupled second-order algebraic equations,
\begin{align}
    -\eta\omega^{2}+\left(\textrm{i}\alpha\pm 1\right)\omega+\Omega_{\pm}(\bm{k})=0,
\end{align}
with
\begin{align}
\Omega_{\pm}(\bm{k})  = & \frac{\gamma}{M}\left(-4J^{zz} + 2 J^{xx} \left[\cos (\bm{k}\cdot \bm{\delta}_1) + \cos (\bm{k}\cdot \bm{\delta}_2)\right]\right. \label{Omegabcc}\nonumber\\
& \left.-2 K^{zz} + MB^{z} \mp 2D \sin (\bm{k}\cdot \bm{\delta}_3)\right),
\end{align}

The dispersion relation for precession and nutation spin waves is given by
\begin{align}
\omega_{\pm\rm Prec}(\bm{k}) &  =  \frac{\pm \left(1-a_{\pm}(\bm{k})\right) +\textrm{i} \alpha\left(1-a^{-1}_{\pm}(\bm{k})\right)}{2\eta}\,,\label{eq:precfreq}\\
\omega_{\pm\rm Nut}(\bm{k}) &  =  \frac{\pm \left(1+a_{\pm}(\bm{k})\right) +\textrm{i} \alpha\left(1+a^{-1}_{\pm}(\bm{k})\right)}{2\eta}\,,\label{eq:nutfreq}
\end{align}
with
\begin{align}
& a_{\pm}(\bm{k})=\nonumber\\
& \sqrt{\frac{\left(1-\alpha^2+4b_{\pm}(\bm{k})\right)+\sqrt{\left(1-\alpha^2+4b_{\pm}(\bm{k})\right)^{2}+4\alpha^{2}}}{2}}\,,
\end{align}
where we introduced the dimensionless factor $b_{\pm}(\bm{k})=\eta \Omega_{\pm}(\bm{k})$. For $a_{\pm}(\bm{k})\ge 1$, the imaginary part of all frequencies is positive, meaning that the spin waves decay over time. It can be shown that this stability condition is satisfied for $\Omega_{\pm}(\bm{k})\ge 0$ for all $\eta\ge 0$ inertial relaxation times. In the following, due to the particle-hole symmetry we only show the solutions with positive real parts, i.e. $\omega_{-\rm Prec}(\bm{k})$ and $\omega_{+\rm Nut}(\bm{k})$ due to $a_{\pm}(\bm{k})\ge 1$.

These frequencies are illustrated in Fig.~\ref{Fig2} for the parameters $J^{zz} = -1.02\times 10^{-21}$ J, $J^{xx} =J^{yy} = -0.99\times 10^{-21}$ J, $D = 10^{-22}$ J, $K^{zz} = - 10^{-22}$ J and $B^{z} = 0$, where the damping was set to $\alpha=0$. From Eqs.~\eqref{eq:precfreq} and \eqref{eq:nutfreq} follows $\omega_{\pm\rm Prec}(\bm{k})+\omega_{\pm\rm Nut}(\bm{k})=\left(\pm 1+\textrm{i}\alpha\right)/\eta$. This means that the relation between the precession and nutation branches in Figs.~\ref{Fig2}(a) and (c) and Figs.~\ref{Fig2}(b) and (d) is given by
\begin{align}
\omega_{+\rm Nut}(\bm{k})=\omega_{-\rm Prec}(-\bm{k})+\eta^{-1},
\end{align}
where $a_{-}(\bm{k})=a_{+}(-\bm{k})$ was used. The two bands are shifted by a constant with respect to each other and inverted in $\boldsymbol{k}$ space. The shift is subtracted from the nutation frequencies for visualization purposes, otherwise the branches would be further away from each other as $\eta$ is varied. The inversion is apparent in Figs.~\ref{Fig2}(c) and (d) where non-reciprocal spin wave propagation $\omega(\bm{k})\neq \omega(-\bm{k})$ can be observed. This is caused by the Dzyaloshinsky--Moriya interaction as can be seen from Eq.~\eqref{Omegabcc}, which creates an asymmetry in the spectrum along the $k_{x}$ direction, dictated by the symmetry of the system. For the selected sign of $D$, nutation spin waves have a lower frequency when propagating along $-k_{x}$, while for precession spin waves have a lower frequency along $k_{x}$. This follows from the fact that at a given lattice site, precession and nutation spin waves have opposite circular polarizations, as it is known for the $\boldsymbol{k}=\boldsymbol{0}$ modes~\cite{Mondal2020nutation}.
For $\alpha = 0$, the two frequencies in Eqs.~\eqref{eq:precfreq} and \eqref{eq:nutfreq} can be expanded in $b_{\pm}(\bm{k})$ as
\begin{align}
\omega_{-\rm Prec}(\bm{k}) 
& \approx \Omega_{-}(\bm{k}) - b_{-}(\bm{k}) \Omega_{-}(\bm{k})\,,\\
\omega_{+\rm Nut}(\bm{k}) & \approx  \frac{1}{\eta} +  \Omega_{+}(\bm{k}) -  b_{+}(\bm{k})\Omega_{+}(\bm{k}).
\end{align}
In this approximation, the original precession frequency $\Omega_{\pm}(\bm{k})$ is renormalized by the factor $1-b_{\pm}(\bm{k})$ due to the inertia. While in the $\boldsymbol{k}=\boldsymbol{0}$ point this is typically small, at high wave vectors $\Omega_{\pm}(\bm{k})$ is dominated by the exchange terms, and $b_{\pm}(\bm{k})$ is orders of magnitude larger. The stronger decrease of the excitation frequencies at higher wave vectors leads to a flattening of the parabolic spin-wave spectrum with increasing inertial relaxation time $\eta$, as shown in Fig.~\ref{Fig2}. 

The most apparent consequence of the inertial dynamics is the appearance of the nutation spin-wave band. Since it is shifted by $1/\eta$ with respect to the precession band, it may be challenging to observe using experimental techniques calibrated to typical precession frequencies. Therefore, it is worthwhile to discuss how the signatures of inertial dynamics show up in the precession band. While the decrease of the spin-wave frequencies between the inertial and non-inertial cases at the edges of the Brillouin zone can achieve relatively high values in this model calculation, it is unlikely that inertial effects can be experimentally demonstrated based on the measurement of the precession spin-wave band alone. The value of the exchange parameters in such measurements is determined based on the observed spectrum, and the flattened dispersion relation in Fig.~\ref{Fig2} may also be fitted by renormalizing the nearest-neighbor exchange and taking interactions with further neighbors into account. The signatures of inertial dynamics could only be proven if the spin interactions obtained from the spectrum are compared to static quantities determined by the spin model parameters but unaffected by the dynamics, such as magnetization curves or the critical temperature.

The group velocity of precession spin waves is given by the expression
\begin{align}
    \boldsymbol{v}_{-{\rm Prec}}\left(\boldsymbol{k}\right) & =  \frac{\textrm{d}\omega_{-\rm Prec}(\bm{k})}{\textrm{d}\boldsymbol{k}} = a^{-1}_{-}(\bm{k})\frac{\textrm{d}\Omega_{-}(\bm{k})}{\textrm{d}\boldsymbol{k}}.\label{groupvel}
\end{align}
\begin{figure}[tbh!]
\centering
\includegraphics[scale =0.25]{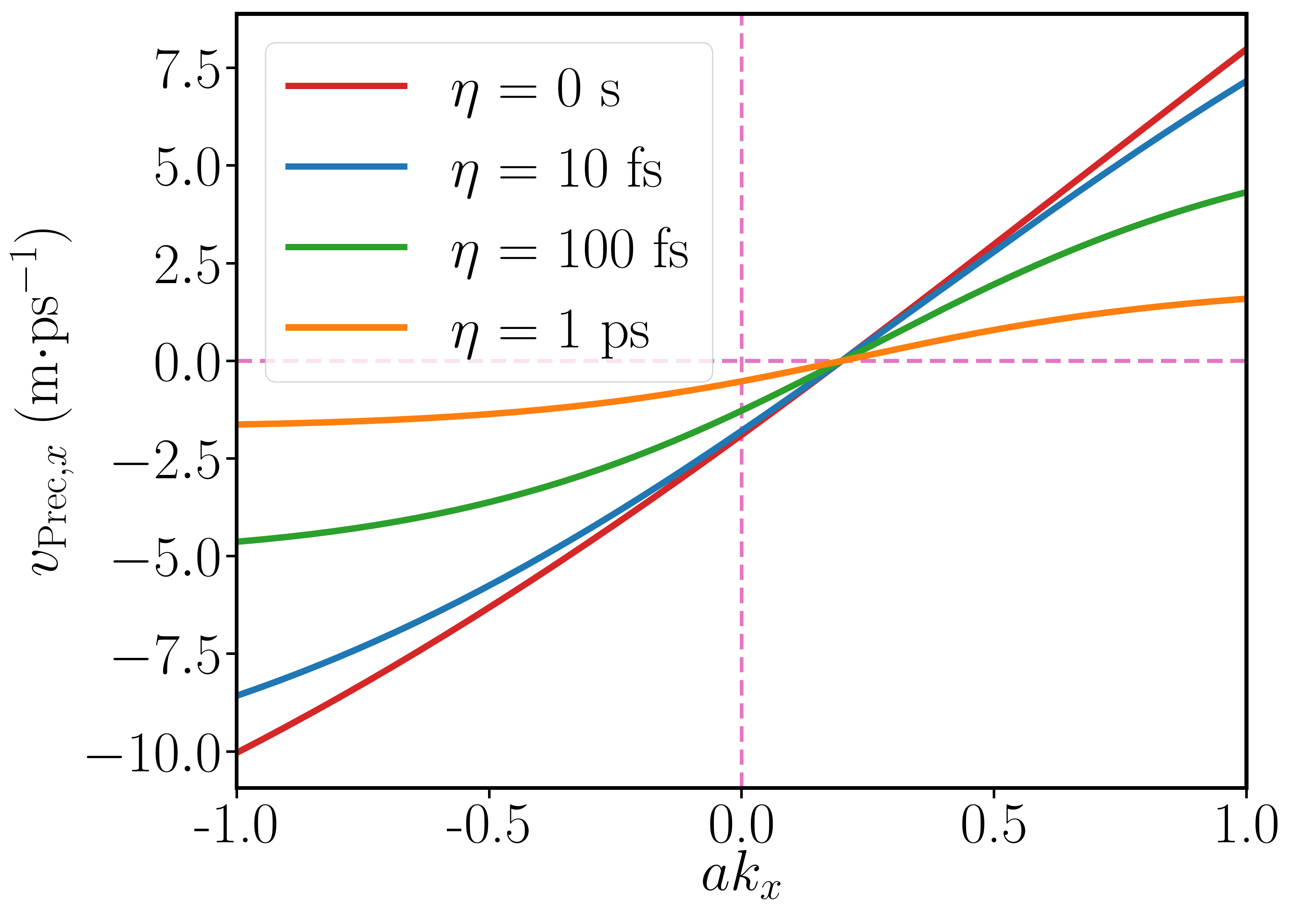}
\caption{The calculated group velocity of precession spin waves at several inertial relaxation times along $k_x$ for $k_z = 0$ with the lattice constant $a = 3$ {\AA}. The parameters are  $J^{zz} = -1.02\times 10^{-21}$ J, $J^{xx} = J^{yy} = -0.99\times 10^{-21}$ J, $D = 10^{-22}$ J, $K^{zz} = -10^{-22}$ J, $B^{z} = 0$ and $\alpha = 0$. 
}
\label{Group_Velocity}
\end{figure}
The group velocity is presented in Fig.\ \ref{Group_Velocity} for the considered model system. Due to the inertial dynamics, $\boldsymbol{v}_{-{\rm Prec}}\left(\boldsymbol{k}\right)$ is rescaled by the factor $a^{-1}_{-}(\bm{k})$. The group velocity vanishes in the minimum of the spin-wave spectrum, which is shifted along the $k_{x}$ direction due to the Dzyaloshinsky--Moriya interaction as discussed above. The position of the minimum is independent of $\eta$. The magnitude of $\boldsymbol{v}_{-{\rm Prec}}$ is reduced for increased inertial relaxation times $\eta$ and further away from the minimum of the spectrum. As it was explained above for the dispersion relation, the experimental determination of the group velocities in only the precession band is probably insufficient for concluding on the strength of inertial effects, since a similar decrease could also be explained by a different set of interaction parameters.   

It is worthwhile to compare the results here to the calculations in Refs.~\cite{Makhfudz2020} and \cite{PhysRevB.104.054425}, to which the formalism here is equivalent if only isotropic exchange interactions are considered, and the Fourier transforms are expanded in $\boldsymbol{k}$ in the long-wavelength limit. 
In Ref.~\cite{Makhfudz2020} it was stated that nutation waves have higher frequencies and speeds compared to conventional precession spin waves. While we can confirm the higher frequencies of the nutation spin waves by the constant shift of $1/\eta$ at a fixed wave vector (the non-reciprocity vanishes in the absence of the Dzyaloshinsky--Moriya interaction) in agreement with Ref.~\cite{PhysRevB.104.054425}, nutation waves have the exact same group velocity given by Eq.~\eqref{Eff_Gyro} as precession waves. A higher speed for the nutation waves can only be observed if different wave vectors are compared between the two bands, as it was performed in Ref.~\cite{Makhfudz2020}.

\begin{figure}[tbh!]
    \centering
    \includegraphics[scale = 0.27]{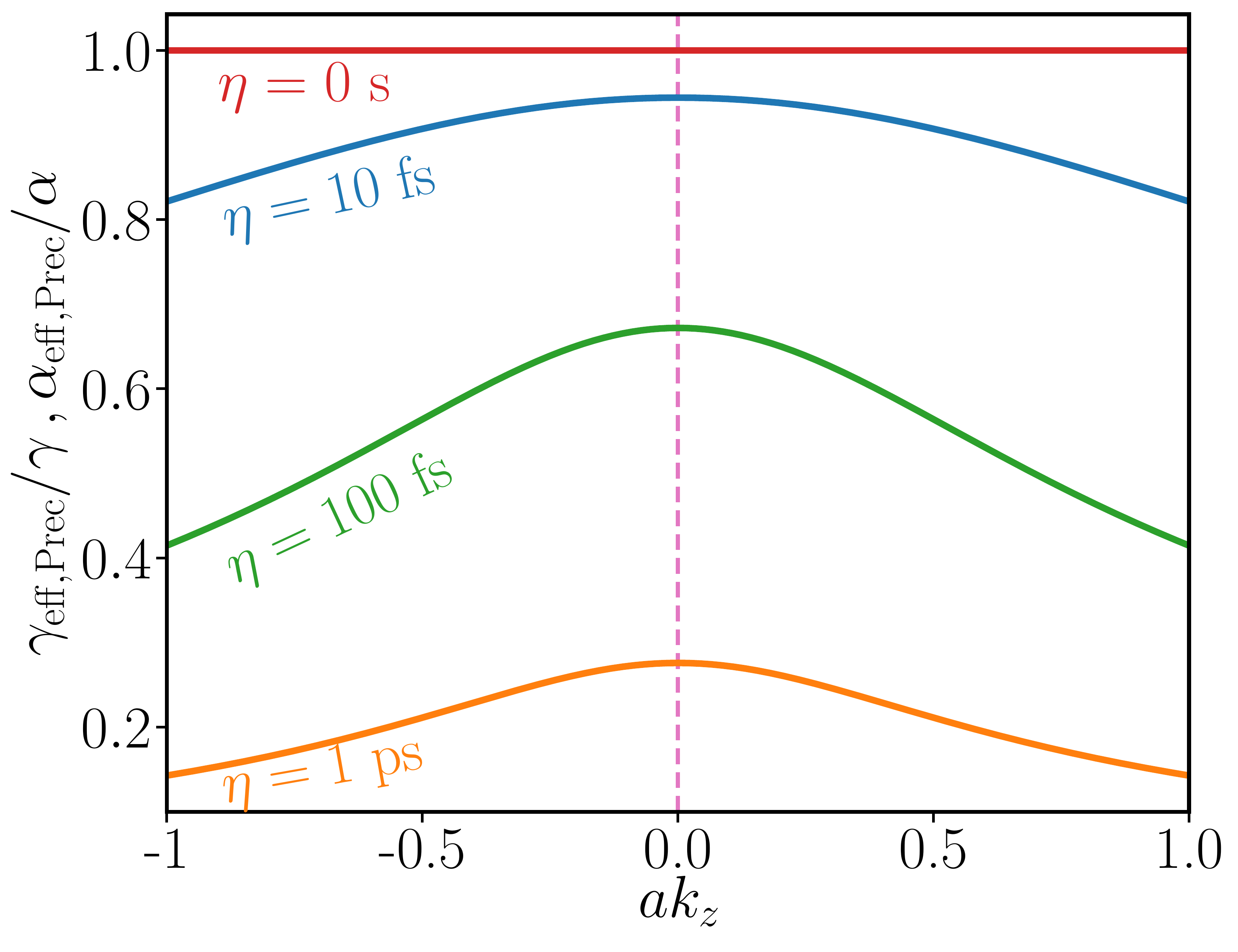}
    \caption{The effective gyromagnetic ratio $\gamma_{\textrm{eff,Prec}}$ and effective damping parameter $\alpha_{\rm eff, Prec}$ from Eq.~\eqref{Eff_Gyro} and Eq.~\eqref{alphaeff} calculated for several inertial relaxation times along $k_z$ for $k_x = 0$. The values are divided by $\gamma$ and $\alpha$ observable for free precession. 
    The parameters are  $J^{zz} = -1.02\times 10^{-21}$ J, $J^{xx} = J^{yy}= -0.99\times 10^{-21}$ J, $D = 10^{-22}$ J, $K^{zz} = -10^{-22}$ J and $B^{z} = 0$.}
    \label{Eff_Gyro_Ratio}
\end{figure}

A closer inspection of the dispersion relation of precession spin waves may shed further light on the consequences of inertial dynamics. For an external magnetic field applied along the $\bm{\hat{z}}$ direction, the effective gyromagnetic ratio is computed as
\begin{align}
    \gamma_{\rm eff}(\bm{k})  = \frac{\textrm{d}\omega_{-\rm Prec}(\bm{k})}{\textrm{d}B^{z}} & = a^{-1}_{-}(\bm{k})\frac{\textrm{d}\Omega_{-}(\bm{k})}{\textrm{d}B^{z}}\nonumber\\
    & =\frac{\gamma}{\sqrt{1+4b_{-}(\bm{k})}}
    \label{Eff_Gyro}
\end{align}
where the last formula holds for $\alpha=0$. This expression is illustrated in Fig.\ \ref{Eff_Gyro_Ratio}. While $\gamma_{\rm eff}$ corresponds to the free gyromagnetic ratio in the absence of inertia, it displays a wave-vector-dependent reduction as $\eta$ increases. The renormalization factor $a^{-1}\left(\bm{k}\right)$ is the same as for the group velocity in Eq.~\eqref{groupvel}. Note that $\gamma_{\rm eff}$ is the same for the nutation band when $a^{-1}_{-}(\bm{k})$ is replaced by $a^{-1}_{+}(\bm{k})$ because of the constant shift and the inversion in $\boldsymbol{k}$ space described above. 

However, a renormalization of the gyromagnetic ratio with the wave vector is not unique to inertial dynamics. It can also be observed in ferromagnets with biaxial anisotropy, see Appendix~\ref{AppendixA} for the calculation; however, the renormalization shows a different qualitative dependence on $\boldsymbol{k}$. In biaxial ferromagnets in the absence of inertia, the gyromagnetic factor is enhanced at short wave vectors and then converges to $\gamma$ on the scale $ka\propto\sqrt{\left(K^{xx}-K^{yy}\right)/J}$, where $K^{xx}-K^{yy}$ is the anisotropy energy between the intermediate and the hard axes. In contrast, in the inertial case the gyromagnetic ratio is always smaller than its free value, and it typically displays a decrease in a broad range in reciprocal space, occurring on the characteristic scale of $J_{\bm{0}}-J_{\bm{k}}\approx\frac{M}{\gamma}\eta$, which is away from the center of the Brillouin zone for typical values of the exchange interaction and of the inertial relaxation time. Therefore, the experimental determination of the gyromagnetic ratio in a wide range of wave vectors should make it possible to distinguish between anisotropic and inertial effects.

The effective damping parameter of the spin waves is defined as
\begin{align}
    \alpha_{\textrm{eff},-\textrm{Prec}}(\bm{k})=\frac{\textrm{Im}\:\omega_{-\rm Prec}(\bm{k})}{\textrm{Re}\:\omega_{-\rm Prec}(\bm{k})}=a_{-}^{-1}(\bm{k})\alpha;\label{alphaeff}
\end{align}
for nutation spin waves $a_{-}^{-1}(\bm{k})$ has to be replaced by $a_{+}^{-1}(\bm{k})$. Since the dependence on the wave vector is precisely the same as for the effective gyromagnetic ratio in Eq.~\eqref{Eff_Gyro}, Fig.~\ref{Eff_Gyro_Ratio} also provides an illustration of the wave-vector-dependence of the effective damping. The analysis also follows similar lines: while the effective damping also depends on the wave vector in biaxial ferromagnets, it is only enhanced in the vicinity of the $\boldsymbol{k}=\boldsymbol{0}$ point, while inertial effects reduce the effective damping and yield a broad variation in the Brillouin zone. Although the Gilbert damping $\alpha$ is material dependent while $\gamma$ is more or less constant between magnetic systems, the different behavior in the Brillouin zone should make it possible to distinguish between inertial and anisotropic effects on the effective damping. Finally, we remark that the enhanced robustness of nutation waves compared to precession waves proposed in Ref.~\cite{Makhfudz2020} only holds if effective damping parameters at different wave vectors are compared. The same value for the effective damping between precession and nutation waves and its decrease at higher wave vectors agrees with the results of Ref.~\cite{PhysRevB.104.054425}, where the quality factor $Q=1/(2\alpha_{\textrm{eff}})$ was calculated.

\section{Antiferromagnets\label{sec3}}
\subsection{Linear spin-wave theory}
Here, we consider two-sublattice antiferromagnets. In the ground state, the magnetic moments on the two sublattices $A$ and $B$ point antiparallel, assumed to lie along the $
+\hat{\boldsymbol{z}}$ and $-\hat{\boldsymbol{z}}$ directions. Introducing the variables 
$\beta_{1i}$, $\beta_{2i}$ on sublattice $i\in A$ and $\beta_{1j}$, $\beta_{2j}$ on sublattice $j\in B$, the spin directions are expanded as
\begin{align}
\label{Eq_anti1}
    \bm{S}_i & = \begin{pmatrix}
    {\beta}_{2i}\\
    - {\beta}_{1i}\\
        1 - \dfrac{{\beta}_{1i}^{2}}{2} - \dfrac{{\beta}_{2i}^{2}}{2}
    \end{pmatrix},\\
    \bm{S}_j & = \begin{pmatrix}
    {\beta}_{2j}\\
     {\beta}_{1j}\\
        - 1 + \dfrac{{\beta}_{1j}^{2}}{2} + \dfrac{{\beta}_{2j}^{2}}{2}
    \end{pmatrix}.
    \label{Eq_anti2}
\end{align}

Following the same procedure as for ferromagnets, in Fourier space and the circularly polarized basis we arrive again at the linearized equation of motion Eq.~\eqref{Eq10}. In this case, the eigenvectors contain four components,
\begin{align}
    \tilde{\boldsymbol{\beta}}\left(\boldsymbol{k}\right)=&\begin{pmatrix}\tilde{\beta}_{+A}\left(\boldsymbol{k}\right) & \tilde{\beta}_{-A}\left(\boldsymbol{k}\right) & \tilde{\beta}_{+B}\left(\boldsymbol{k}\right) & \tilde{\beta}_{-B}\left(\boldsymbol{k}\right)\end{pmatrix}^{T},\\
    D\left(\omega\right)=&\textrm{diag}\left(D_{+A}\left(\omega\right) , D_{-A}\left(\omega\right) , D_{+B}\left(\omega\right) , D_{-B}\left(\omega\right)\right),
\end{align}
with
\begin{align}
     D_{\pm A/B}= &-\eta_{A/B}\omega^{2}+\left(\textrm{i}\alpha_{A/B} \pm 1 \right)\omega.
\end{align}
The derivation and the expression for $H_{\textrm{SW}}\left(\bm{k}\right)$ are given in Appendix~\ref{AppendixB}.

\begin{figure}[tbh!]
    \centering
    \includegraphics[scale = 0.44]{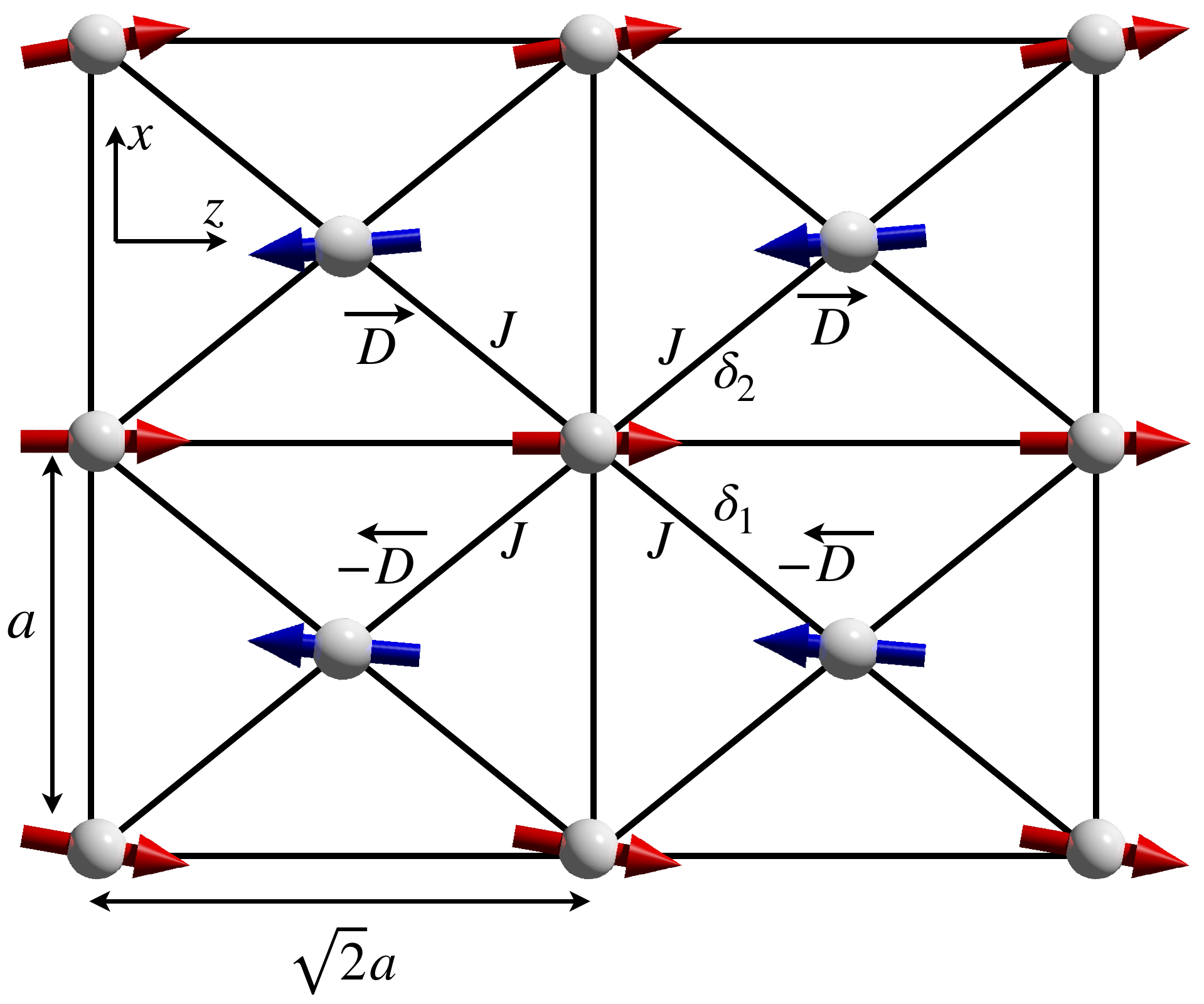}
    \caption{Sketch of a two-dimensional centered rectangular lattice with nearest-neighbor antiferromagnetic coupling. The lattice constants are $\sqrt{2}a$ and $a$ along the $z$ and $x$ directions, respectively. The arrows show the ground state spin orientation along the equilibrium direction $z$, with a small tilting of spins illustrating the traveling spin wave. The nearest-neighbor lattice vectors are labeled as $\boldsymbol{\delta}_1$ and $\boldsymbol{\delta}_2$. The nearest-neighbor coupling coefficients are denoted by $J$ with $J^{xx}=J^{yy}$ and $J^{zz}$, while the nearest-neighbor Dzyaloshinsky-Moriya vectors are $\boldsymbol{D}$.}
    \label{Sketch_Antiferro}
\end{figure} 

The equation of motion stil satisfies particle-hole symmetry, implying $\omega\left(-\boldsymbol{k}\right)=-\omega^{*}\left(\boldsymbol{k}\right)$ for the eigenvalues and $\tilde{\boldsymbol{\beta}}\left(-\boldsymbol{k}\right)=\mathcal{C}\tilde{\boldsymbol{\beta}}\left(\boldsymbol{k}\right)$ for the eigenvectors. Now we consider the antiferromagnetic limit, where the two sublattices are equivalent, i.e., $\gamma_{A}=\gamma_{B} = \gamma$, $\alpha_{A}=\alpha_{B} = \alpha$, $\eta_{A}=\eta_{B} = \eta$, $M_{A}=M_{B}$, $\boldsymbol{K}_{A}=\boldsymbol{K}_{B}$, and $\boldsymbol{J}_{\boldsymbol{k}AB}=\boldsymbol{J}_{\boldsymbol{k}BA}$. The external magnetic field is set to zero ($B^{z}=0$), and the $A$ and $B$ sublattices are assumed to form a Bravais lattice together. In this case and for $\alpha=0$, the eigenvalues $\omega\left(-\boldsymbol{k}\right)=\omega\left(\boldsymbol{k}\right)$ and the eigenvectors  $\tilde{\boldsymbol{\beta}}\left(-\boldsymbol{k}\right)=\mathcal{T}\tilde{\boldsymbol{\beta}}\left(\boldsymbol{k}\right)$ are divided into pairs between wave vectors $\boldsymbol{k}$ and $-\boldsymbol{k}$ by the effective time-reversal symmetry $\mathcal{T}=\tau^{x}\mathcal{K}$, where $\tau^{x}$ is a Pauli matrix exchanging the two sublattices. $\mathcal{T}$ satisfies certain natural requirements on time reversal, namely that it is broken if the system has a finite net magnetic moment due to the inequivalence of the sublattices or because of the application of an external field, and it does not hold in the presence of dissipation $\alpha$. 

\subsection{Dispersion relation}
\begin{figure*}[tbh!]
    \centering
    \includegraphics[scale = 0.26]{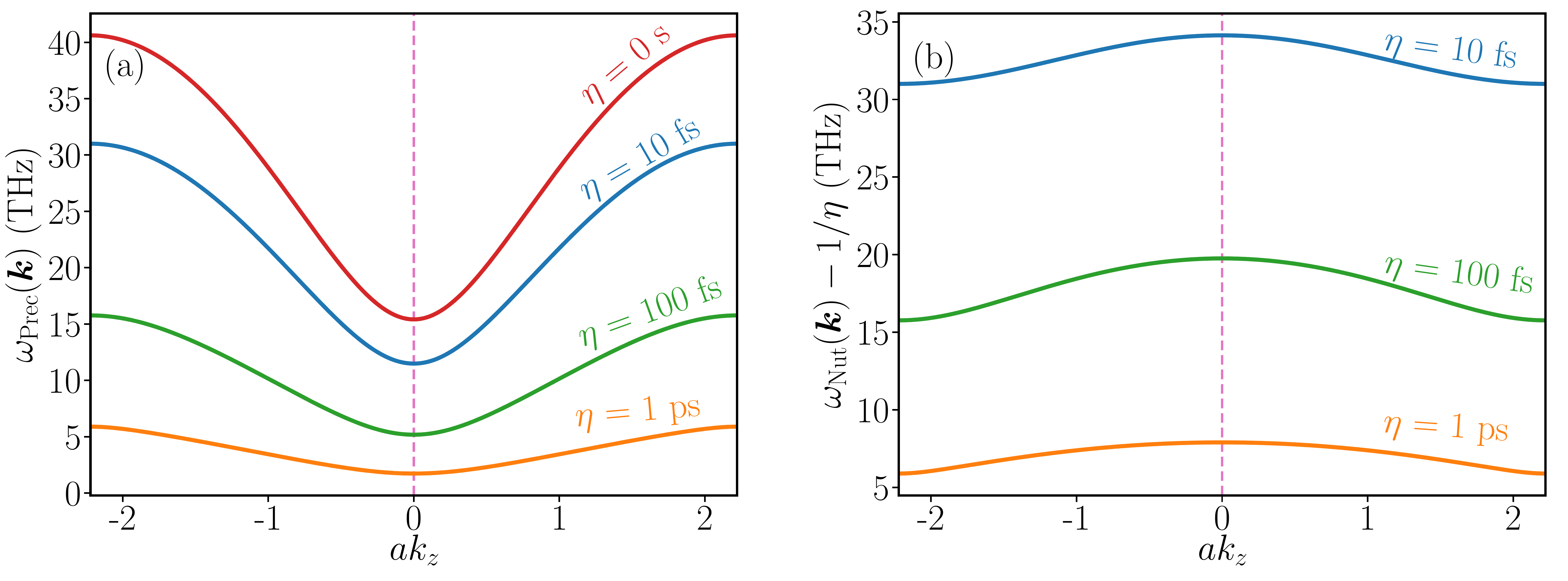}
    \includegraphics[scale = 0.26]{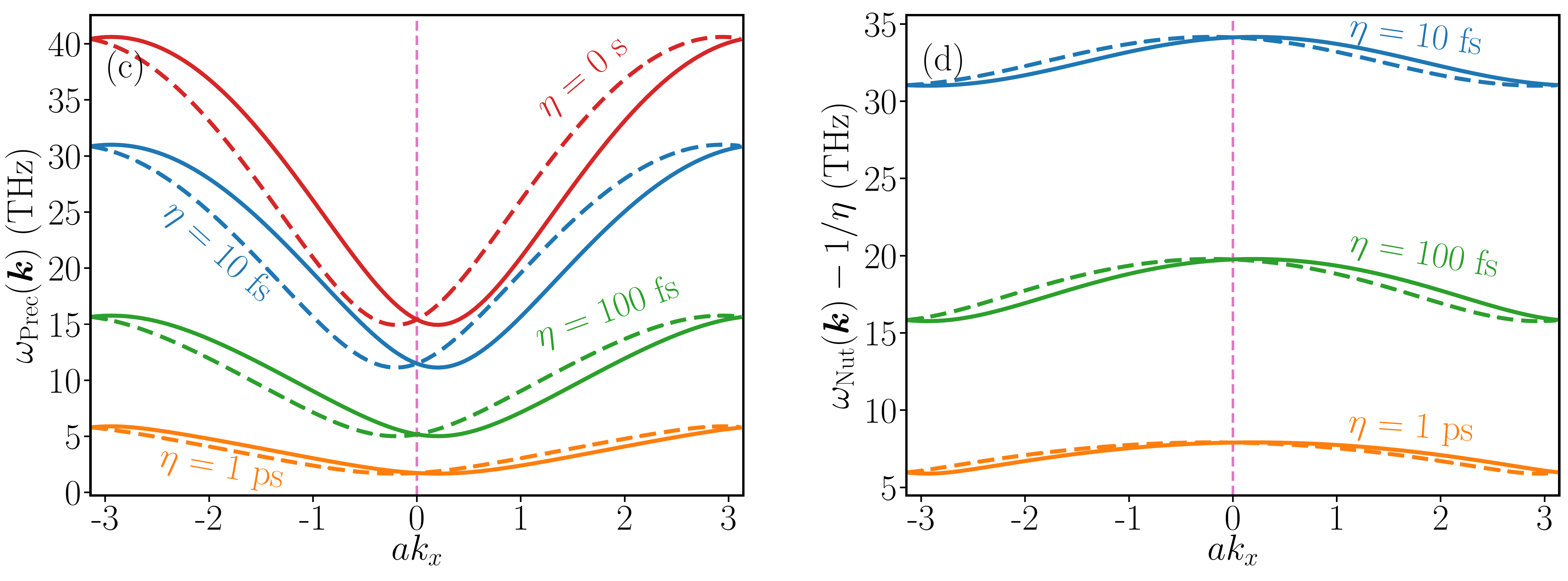}
    \caption{Spin-wave dispersion relation in the antiferromagnet without and with the inertia, computed based on Eq.~\eqref{Secular_AFM}. Solid and dashed lines indicate the $+$ and $-$ branches of the dispersion relation. (a) Precession and  (b) nutation bands along $k_z$ for $k_x = 0$. (c) Precession and  (d) nutation bands along $k_x$ for $k_z = 0$. The parameters are  $J^{zz} = 1.02\times 10^{-21}$ J, $J^{xx} = 0.99\times 10^{-21}$ J, $D = -10^{-22}$ J, $K^{zz} = -10^{-22}$ J, $B^{z} = 0$ and $\alpha = 0$.  
    }
    \label{AFM_Freq}
\end{figure*}

For illustration, we again consider the centered rectangular lattice sketched in Fig.~\ref{Sketch_Antiferro}. Such a local antiferromagnetic order have been observed in monolayers of Mn~\cite{Bode2007} and Cr~\cite{Santos2008} on W(110), although on longer length scales they are modulated to form spin spirals. 
We only consider exchange ($J^{zz}, J^{xx}=J^{yy}$) and Dzyaloshinsky--Moriya ($D_{ij}=J_{ij}^{xy}=-J_{ij}^{yx}$) interactions between the nearest neighbors belonging to different sublattices, and a uniaxial on-site anisotropy term $K_{A/B}^{zz}$. 
In this model, the linearized equations of motion decouple into two two-by-two systems of equations containing the components $\left(\tilde{\beta}_{+A}\left(\boldsymbol{k}\right),\tilde{\beta}_{-B}\left(\boldsymbol{k}\right)\right)$ and $\left(\tilde{\beta}_{-A}\left(\boldsymbol{k}\right),\tilde{\beta}_{+B}\left(\boldsymbol{k}\right)\right)$, respectively. Denoting the first pair by $+$ and the second pair by $-$, we obtain the following secular equation for the eigenfrequencies:    
 \begin{align}
    \mathcal{A}_{\pm} \omega^4(\bm{k}) + \mathcal{B}_{\pm}\omega^3(\bm{k}) + \mathcal{C}_{\pm}\omega^2(\bm{k})
     + \mathcal{D}_{\pm} \omega(\bm{k}) + \mathcal{E}_{\pm}(\bm{k}) = 0,
     \label{Secular_AFM}
\end{align}
where the following parameters have been defined:
\begin{align}
    \mathcal{A}_{\pm}  = &\eta_A\eta_B,\\
    \mathcal{B}_{\pm}  = &\left[\pm(\eta_A-\eta_B) - {\rm i} \left(\alpha_B\eta_A +\alpha_A\eta_B \right)\right],\\
    \mathcal{C}_{\pm}  = &\left[-1 -\alpha_A\alpha_B - (\Omega_{AA}\eta_B +\Omega_{BB}\eta_A ) \mp {\rm i} (\alpha_A - \alpha_B)\right],\\
    \mathcal{D}_{\pm}  = &\left[\mp\left(\Omega_{AA} - \Omega_{BB}\right) + {\rm i}\left(\Omega_{AA}\alpha_B + \Omega_{BB}\alpha_A\right)\right],\\
    \mathcal{E}_{\pm}(\bm{k})  = &\Omega_{AA}\Omega_{BB} - W_{\pm AB}(\bm{k})W_{\mp BA}(\bm{k}),\\
    \Omega_{AA} =& \frac{\gamma_{A}}{M_{A}}\left(4J^{zz} -2 K_{A}^{zz} + M_{A} B^{z}\right),\\
    \Omega_{BB} =& \frac{\gamma_{B}}{M_{B}}\left(4J^{zz} -2 K_{B}^{zz} -M_{B}B^{z}\right),\\
    W_{\pm AB}(\bm{k})= & \frac{\gamma_{A}}{M_{A}}\left(2 J^{xx} \left[\cos (\bm{k}\cdot \bm{\delta}_1) + \cos (\bm{k}\cdot \bm{\delta}_2)\right]\right. \nonumber\\ &\left.\mp 2D \left[\sin (\bm{k}\cdot \bm{\delta}_2) - \sin (\bm{k}\cdot \bm{\delta}_1)\right]\right),\\
    W_{\pm BA}(\bm{k})= & \frac{\gamma_{B}}{M_{B}}\left(2 J^{xx} \left[\cos (\bm{k}\cdot \bm{\delta}_1) + \cos (\bm{k}\cdot \bm{\delta}_2)\right] \right. \nonumber\\ &\left.\pm 2D \left[\sin (\bm{k}\cdot \bm{\delta}_2) - \sin (\bm{k}\cdot \bm{\delta}_1)\right]\right).
\end{align}

We present the analytical solution of Eq.~\eqref{Secular_AFM} only for two identical sublattices, and set the damping and the external field to zero. This leads to the simplification $\Omega=\Omega_{AA}=\Omega_{BB}$ and $W_{\pm}=W_{\pm AB}=W_{\mp BA}$. The dispersion relation is given by
\begin{align}
    \omega_{\pm \textrm{Prec}}^{2}\left(\boldsymbol{k}\right)=&\frac{1+2b}{2\eta^{2}}\left(1-\sqrt{1-\frac{4\eta^{2}}{\left(1+2b\right)^2}\left(\Omega^{2}-W^{2}_{\pm}\left(\boldsymbol{k}\right)\right)}\right),\\
    \omega_{\pm \textrm{Nut}}^{2}\left(\boldsymbol{k}\right)=&\frac{1+2b}{2\eta^{2}}\left(1+\sqrt{1-\frac{4\eta^{2}}{\left(1+2b\right)^2}\left(\Omega^{2}-W^{2}_{\pm}\left(\boldsymbol{k}\right)\right)}\right)
\end{align}
for precession and nutation spin waves, respectively. Here $b=\eta\Omega$ was introduced, analogously to $b_{\pm}\left(\bm{k}\right)$ in the ferromagnetic case. We will only consider the $\omega_{\pm \textrm{Prec,Nut}}\left(\boldsymbol{k}\right)\ge 0$ solutions due to particle-hole symmetry.

The spin-wave frequencies are shown in Fig.~\ref{AFM_Freq} for two equivalent sublattices with the parameters $J^{zz} = 1.02\times 10^{-21}$ J, $J^{xx} =J^{yy} = 0.99\times 10^{-21}$ J, $D = -10^{-22}$ J, $K^{zz} = - 10^{-22}$ J and in the absence of damping. In Fig.~\ref{AFM_Freq}(c) and (d) it can be seen that the individual branches $\omega_{+ \textrm{Prec}}\left(\boldsymbol{k}\right),\omega_{- \textrm{Prec}}\left(\boldsymbol{k}\right),\omega_{+ \textrm{Nut}}\left(\boldsymbol{k}\right)$, and $\omega_{- \textrm{Nut}}\left(\boldsymbol{k}\right)$ are non-reciprocal due to the presence of the Dzyaloshinsky--Moriya interaction. However, the branches are connected by the reciprocal symmetry $\mathcal{R}=\tau^{x}$, ensuring $\omega_{+ \textrm{Prec}}\left(\boldsymbol{k}\right)=\omega_{- \textrm{Prec}}\left(-\boldsymbol{k}\right)$ and $\omega_{+ \textrm{Nut}}\left(\boldsymbol{k}\right)=\omega_{- \textrm{Nut}}\left(-\boldsymbol{k}\right)$. In contrast to the time-reversal symmetry $\mathcal{T}=\mathcal{R}\mathcal{K}$ introduced above, reciprocal symmetry requires certain spatial symmetries beyond the equivalence of the two sublattices. These are satisfied by the present system as discussed in Appendix~\ref{AppendixA}. Note that reciprocal symmetry also holds for finite damping $\alpha$, unlike time-reversal symmetry. 

While in the ferromagnetic case the sum of the precession and nutation frequencies was fixed in the same spin-wave branch while $\omega_{+\textrm{Prec}}\left(\boldsymbol{k}\right)$ and $\omega_{+\textrm{Nut}}\left(\boldsymbol{k}\right)$ had opposite signs, in the antiferromagnetic case the sum of the squared frequencies is a constant as a function of wave vector,
\begin{align}
    \omega^{2}_{\pm\textrm{Prec}}\left(\boldsymbol{k}\right)+\omega^{2}_{\pm\textrm{Nut}}\left(\boldsymbol{k}\right)=\frac{1+2b}{\eta^{2}}.
\end{align}
This leads to a considerably different spectrum. While in the ferromagnet the precession and nutation bands are only shifted with respect to each other and inverted in $\boldsymbol{k}$ space, in the antiferromagnet the nutation frequencies are smaller for higher precession frequencies, meaning that they decrease when moving away from the center of the Brillouin zone. This can also be seen when expanding the frequencies in the parameter $b$, yielding
\begin{align}
\omega_{\pm \textrm{Prec}}\left(\boldsymbol{k}\right)\approx &\sqrt{\frac{\Omega^{2}-W^{2}_{\pm}\left(\boldsymbol{k}\right)}{1+2b}},\\
\omega_{\pm \textrm{Nut}}\left(\boldsymbol{k}\right)\approx &\frac{\sqrt{1+2b}}{\eta}-\frac{\eta\left(\Omega^{2}-W^{2}_{\pm}\left(\boldsymbol{k}\right)\right)}{2\left(1+2b\right)^{\frac{3}{2}}}.
\end{align}
While the ordinary precession frequencies $\omega_{\pm}^{(0)}\left(\boldsymbol{k}\right)=\sqrt{\Omega^{2}-W^{2}_{\pm}\left(\boldsymbol{k}\right)}$ are decreased by a factor of $\left(1+2b\right)^{-\frac{1}{2}}$, the nutation frequency is around $\sqrt{1+2b}/\eta$, which is enhanced compared to the ferromagnetic case by $\sqrt{1+2b}$, as it was already discussed in Ref.~\cite{Mondal2020nutation} in the $\boldsymbol{k}=\boldsymbol{0}$ case. While the precession band is almost linear in wave vector except very close to the minimum where the anisotropy leads to a quadratic dispersion, the nutation band behaves as $\propto -\eta\left(\omega_{\pm}^{(0)}\left(\boldsymbol{k}\right)\right)^{2}$ at low wave vectors, which describes a broad quadratic dispersion with a small negative curvature, as is also visible in Fig.~\ref{AFM_Freq}(b) and (d). Note that the nutation bands are shifted down by $1/\eta$ in the figure as in the ferromagnetic case, which here does not lead them to coincide with the precession bands, but it does make it possible to visualize them in the same frequency range regardless of the value of $\eta$.

\begin{figure}
    \centering
    \includegraphics[scale = 0.25]{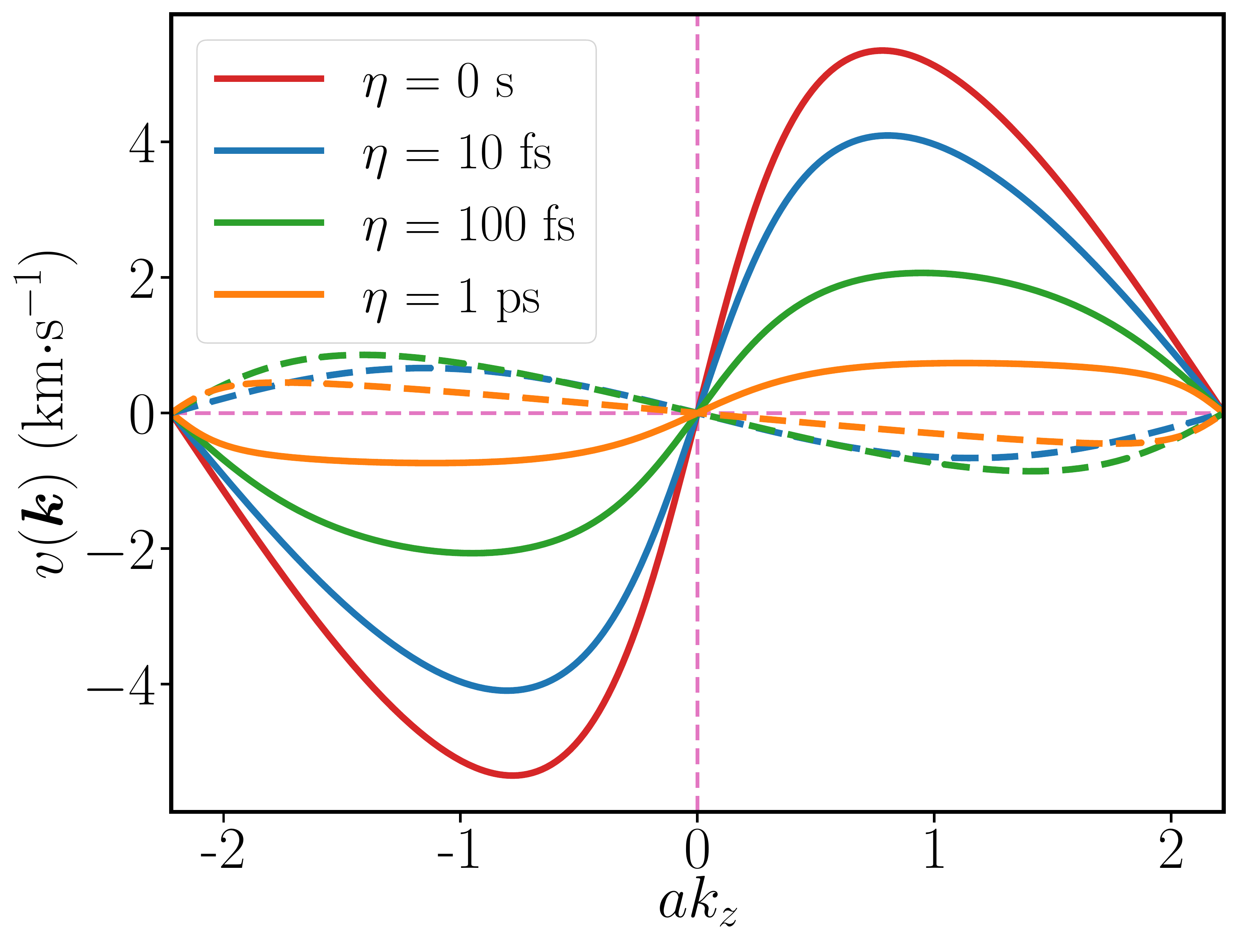}
    \caption{Group velocity of spin waves in antiferromagnets along $k_z$ for $k_x = 0$, computed based on Eq.~\eqref{Secular_AFM}. The solid and dashed lines are the group velocities for precession waves and nutation waves, respectively. The used parameters are  $J^{zz} = 1.02\times 10^{-21}$ J, $J^{xx} = 0.99\times 10^{-21}$ J, $D = -10^{-22}$ J, $K^{zz} = -10^{-22}$ J, $B^z = 0$, $\alpha = 0$, and lattice constant $a = 3$ {\AA}.}
    \label{fig:group_velocity}
\end{figure}

The group velocity in the antiferromagnetic case is
\begin{align}
    &\boldsymbol{v}_{\pm \textrm{Prec}}\left(\boldsymbol{k}\right)\nonumber\\
    \approx & \frac{1}{\sqrt{1-\frac{4\eta^{2}}{\left(1+2b\right)^2}\left(\Omega^{2}-W^{2}_{\pm}\right)}}\frac{1}{\sqrt{1+2b}}\boldsymbol{v}^{(0)}_{\pm \textrm{Prec}}\left(\boldsymbol{k}\right),\label{groupvel_AFM_Prec}\\
   & \boldsymbol{v}_{\pm \textrm{Nut}}\left(\boldsymbol{k}\right)\nonumber\\
   \approx & -\frac{1}{\sqrt{1-\frac{4\eta^{2}}{\left(1+2b\right)^2}\left(\Omega^{2}-W^{2}_{\pm}\right)}}\frac{2\omega_{\pm}^{(0)}\left(\boldsymbol{k}\right)\eta}{\left(1+2b\right)^{\frac{3}{2}}}\boldsymbol{v}^{(0)}_{\pm \textrm{Prec}}\left(\boldsymbol{k}\right),\label{groupvel_AFM_Nut}
\end{align}
where $\boldsymbol{v}^{(0)}_{\pm \textrm{Prec}}\left(\boldsymbol{k}\right)=\frac{1}{2\omega_{\pm}^{(0)}\left(\boldsymbol{k}\right)}\frac{\partial \left(\omega_{\pm}^{(0)}\left(\boldsymbol{k}\right)\right)^{2}}{\partial \boldsymbol{k}}$ is the group velocity in the absence of inertia.

The group velocity is shown in Fig.~\ref{fig:group_velocity} along the $k_{z}$ direction. Compared to the ferromagnetic case in Fig.~\ref{Group_Velocity}, the slope of the group velocity is higher at low wave vector, indicative of the high curvature of the dispersion relation. The leading correction of $\boldsymbol{v}_{\pm \textrm{Prec}}\left(\boldsymbol{k}\right)$ due to the inertial dynamics is a renormalization by a factor of $\left(1+2b\right)^{-\frac{1}{2}}$, similarly to the frequency itself. Since this renormalization is independent of the wave vector, it would not be possible to detect it experimentally, unless the interaction parameters are known from independent measurements as mentioned in the ferromagnetic case. The first term on the right-hand side of Eq.~\eqref{groupvel_AFM_Prec} describes a further decrease of the group velocity, which only becomes relevant at high wave vectors where $\eta^{2}\left(\omega_{\pm}^{(0)}\left(\boldsymbol{k}\right)\right)^{2}$ is relatively large. The group velocity of nutation spin waves points along the opposite direction compared to precession spin waves, and here it is considerably lower in magnitude for all wave vectors because of the flat dispersion relation.

\begin{figure}[tbh!]
    \centering
    \includegraphics[scale = 0.23]{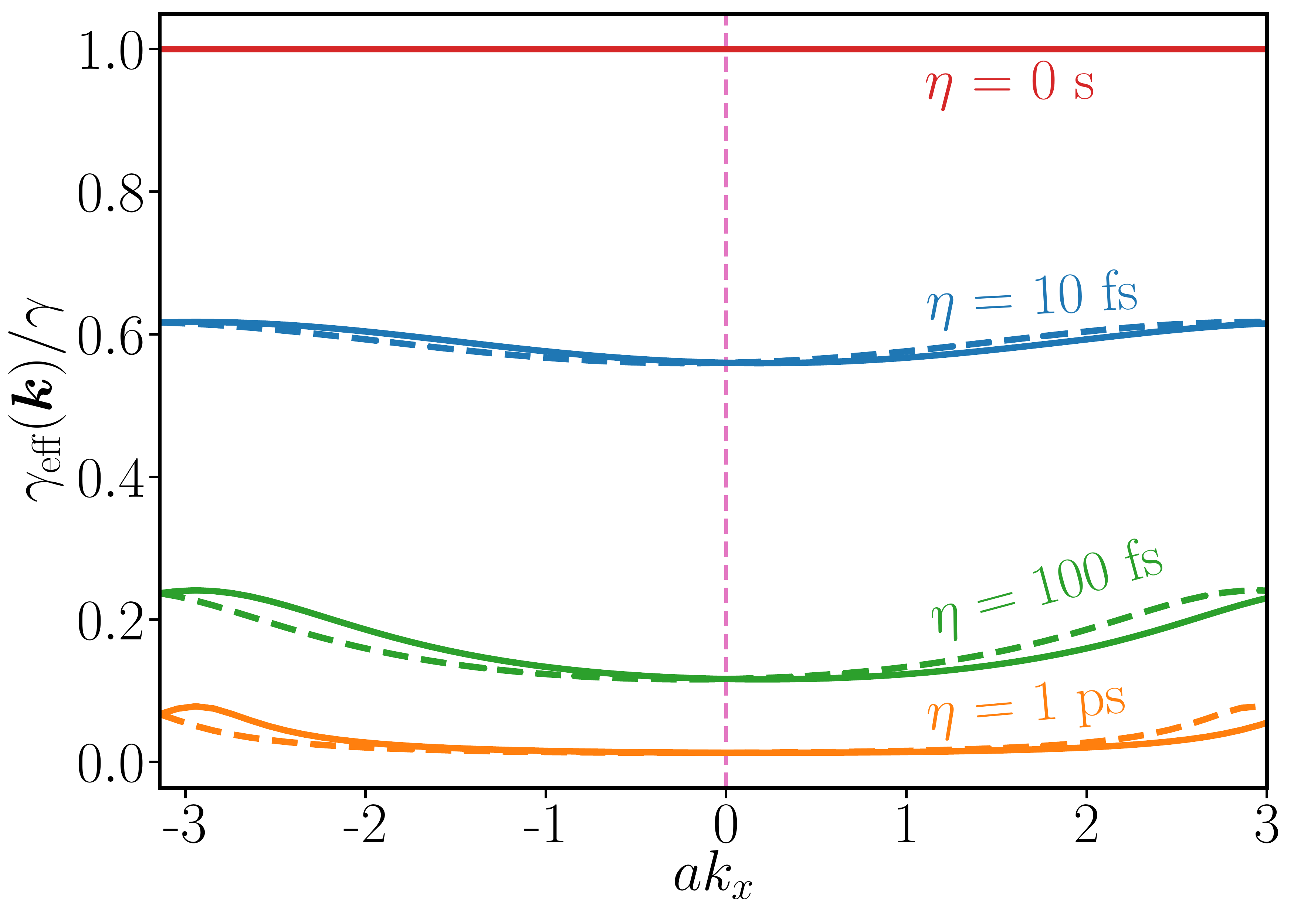}
    \caption{Effective gyromagnetic ratio for spin waves in antiferromagnets along $k_x$ for $k_z = 0$, computed based on Eq.~\eqref{Secular_AFM}. The solid and dashed lines are the gyromagnetic ratios for precession waves and nutation waves, respectively. 
    The used parameters are  $J^{zz} = 1.02\times 10^{-21}$~J, $J^{xx} = 0.99\times 10^{-21}$~J, $D = -10^{-22}$~J, $K^{zz} = -10^{-22}$~J, and $\alpha = 0$. 
    }
    \label{Gyro_Eff_Antiferro}
\end{figure}

Similarly to the ferromagnetic case, deriving the effective gyromagnetic ratios and damping parameters from the dispersion relation could provide signatures for the experimental detection of the inertial dynamics if only the frequency range of the precession spin waves is accessible. The gyromagnetic ratios are expressed as
\begin{align}
    & \gamma_{\textrm{eff,Prec}\pm}\left(\boldsymbol{k}\right)=-\gamma_{\textrm{eff,Nut}\pm}\left(\boldsymbol{k}\right)\approx\nonumber\\
    & \pm\frac{1}{1+2b}\left(1+\frac{2\eta^{2}\left(\omega_{\pm}^{(0)}\left(\boldsymbol{k}\right)\right)^{2}}{\left(1+2b\right)^2}\right)\gamma,\label{gamma_AFM}
\end{align}
while the effective damping parameters read
\begin{align}
\alpha_{\textrm{eff,Prec}}\left(\boldsymbol{k}\right)\approx&\frac{\Omega-\frac{\eta\left(\omega_{\pm}^{(0)}\left(\boldsymbol{k}\right)\right)^{2} }{\left(1+2b\right)}}{\omega_{\pm}^{(0)}\left(\boldsymbol{k}\right)\left(1+2b\right)^{\frac{1}{2}}}\alpha,\label{alpha_AFM_Prec}\\
    \alpha_{\textrm{eff,Nut}}\left(\boldsymbol{k}\right)\approx&\left[\frac{1+b}{\left(1+2b\right)^{\frac{3}{2}}}+\frac{3-5b}{2\left(1+2b\right)^{\frac{7}{2}}}\eta^{2}\left(\omega_{\pm}^{(0)}\left(\boldsymbol{k}\right)\right)^{2}\right]\alpha;\label{alpha_AFM_Nut}
\end{align}
see Appendix~\ref{AppendixB} for the idea of the derivation. These expressions are linearized in the small parameter $\eta^{2}\left(\omega_{\pm}^{(0)}\left(\boldsymbol{k}\right)\right)^{2}\ll 1$, which should be valid at low wave vectors for typical interaction and inertial parameters.

\begin{figure}[tbh!]
    \centering
    \includegraphics[scale = 0.23]{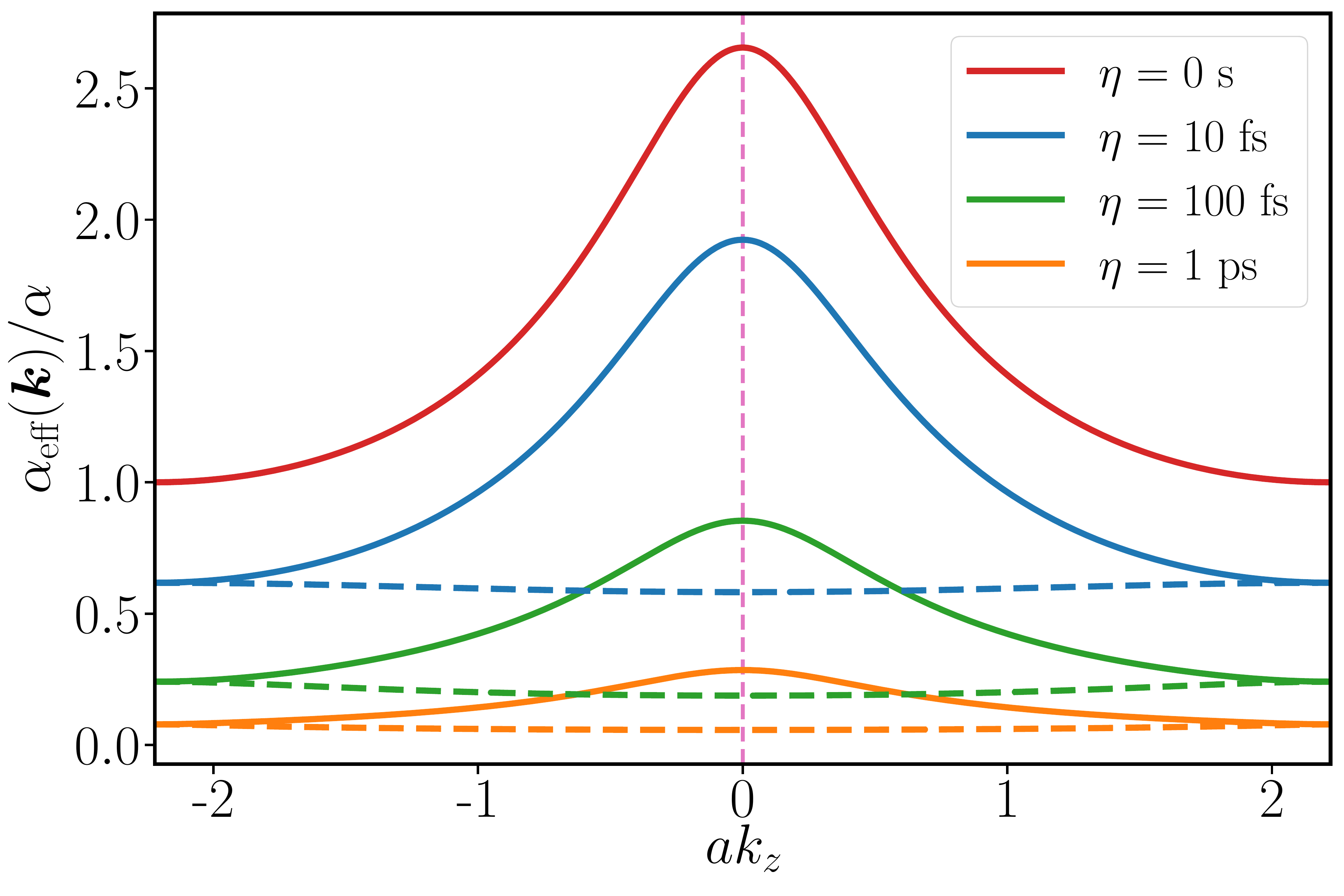}
    \caption{Effective damping for spin waves in antiferromagnets along $k_z$ for $k_x = 0$, computed based on Eq.~\eqref{Secular_AFM}. The solid and dashed lines represent effective damping for precession and nutation modes, respectively. The used parameters are  $J^{zz} = 1.02\times 10^{-21}$ J, $J^{xx} = 0.99\times 10^{-21}$ J, $D = -10^{-22}$ J, $K^{zz} = -10^{-22}$ J, $B^{z} = 0$, and $\alpha = 0.05$.}
    \label{Damping_Antiferro}
\end{figure}

The gyromagnetic ratios are illustrated in Fig.~\ref{Gyro_Eff_Antiferro}. Note that the two spin-wave branches $\omega_{\pm}$ are described by opposite signs of the gyromagnetic ratio, i.e., one is increasing in external magnetic field while the other is decreasing. This is a consequence of the antiferromagnetic alignment of the spins irrespective of the inertial dynamics. Only the positive gyromagnetic ratio is shown in Fig.~\ref{Gyro_Eff_Antiferro}. Due to the finite value of $\eta$, the effective gyromagnetic ratio of precession spin waves in Eq.~\eqref{gamma_AFM} is decreased at zero wave vector stronger than in ferromagnets, since $b_{\textrm{AFM}}\gg b_{\textrm{FM}}\left(\bm{k} = \bm{0}\right)$. However, at increasing wave vector the gyromagnetic ratio increases, in contrast to the ferromagnetic case. The relative change in $\gamma_{\textrm{eff,Prec}}$ between the center and the boundary of the Brillouin zone is smaller in antiferromagnets than in ferromagnets due to a decreased prefactor.  
While the dependence on the wave vector is caused entirely by inertial effects in this model, a biaxial anisotropy may enhance the gyromagnetic ratio at low wave vectors even without inertia, as already discussed for the ferromagnetic case. According to Eq.~\eqref{gamma_AFM}, the gyromagnetic ratio for the nutation band is inverted in $\boldsymbol{k}$ compared to the precession band, when the branches $\gamma_{\textrm{eff,Prec}+}\left(\boldsymbol{k}\right)$ and $\gamma_{\textrm{eff,Prec}-}\left(\boldsymbol{k}\right)$ with positive gyromagnetic ratios are considered, as is also shown in Fig.~\ref{Gyro_Eff_Antiferro}. 

The effective damping parameters are shown in Fig.~\ref{Damping_Antiferro}. Already in the absence of inertia, $\alpha_{\textrm{eff}}$ is increased by a factor of $\Omega/\omega_{\pm}^{(0)}\left(\boldsymbol{k}\right)$ with respect to $\alpha$, which constitutes an exchange enhancement $\propto\sqrt{J/K}$ at low wave vectors and decreases away from the center of the Brillouin zone. This effect is qualitatively similar to the case of biaxial ferromagnets mentioned above. The leading correction due to inertial dynamics is once again the renormalization by the constant factor $\left(1+2b\right)^{-\frac{1}{2}}$, decreasing the effective damping. However, the renormalization by a factor independent of the wave vector is not possible to detect experimentally if the value of the Gilbert damping is not known in advance. Expanding the effective damping in $\eta$ leads to a sub-leading correction which leads to a stronger decrease away from the center of the Brillouin zone; however, based on Fig.~\ref{Damping_Antiferro} this correction may be difficult to discern since the shape is already qualitatively similar without inertia.  
Nutation spin waves are not influenced by the exchange enhancement~\cite{Mondal2020nutation}, and their effective damping parameters are reduced compared to the Gilbert damping $\alpha$ by a factor of $\left(1+b\right)/\left(1+2b\right)^{\frac{3}{2}}$. The dependence on the wave vector only appears when expanding up to second order in $\eta$, leading to a weak increase at higher wave vectors if $b$ is sufficiently small.

\section{Conclusions}

We calculated the spin-wave dispersion relation in ferromagnets and antiferromagnets in the linear spin-wave approximation in the presence of an inertial term in the equation of motion. The inertial term doubles the number of spin-wave bands due to the appearance of nutation spin waves at high frequency, and also decreases the conventional precession spin-wave frequencies due to the hybridization between the bands. The decrease of the precession spin-wave frequencies becomes more pronounced at higher wave vectors. In uniaxial ferromagnets, the nutation band is shifted by $\eta^{-1}$ compared to the precession band, and is inverted in reciprocal space. If the spin-wave propagation is non-reciprocal, for example due to the presence of the Dzyaloshinsky--Moriya interaction, then the preferential direction for spin-wave propagation is inverted between precession and nutation spin waves. In antiferromagnets with two identical sublattices, the sum of the squares of the precession and nutation spin waves is independent of the wave vector. This leads to the fact that the dispersion relation of the nutation waves is quadratic at low wave vectors in contrast to the typical linear dispersion of antiferromagnetic precession spin waves, and the frequency of nutation spin waves decreases at higher wave vectors. The group velocity of spin waves is reduced as a consequence of inertial effects, and nutation spin waves in antiferromagnets propagate in the opposite direction compared to their wave vectors.

While the direct detection of nutation spin waves would provide the most direct evidence of inertial effects on spin dynamics, analyzing the precession band alone may also provide indications for inertial motion. Since the renormalization of spin-wave frequencies and the group velocity may also be explained by a different set of spin-model parameters, these parameters could be fixed by the measurement of static properties such as the critical temperature which is not affected by the inertia. Alternatively, the decrease of the effective gyromagnetic ratio or the effective damping parameter with increasing wave vector could hint towards the role of inertial dynamics. Although a similar wave-vector-dependence may also be caused by anisotropy terms, those modify the gyromagnetic ratio and the effective damping typically at lower frequencies compared to where inertial effects are the most pronounced. Such signatures are expected to be more pronounced in ferromagnets than in antiferromagnets, since in the latter the frequencies of precession spin waves and all derived quantities are renormalized by a factor which depends very weakly on the wave vector. Such a constant renormalization cannot be determined from the investigation of spin waves unless the interaction parameters, the gyromagnetic ratio or the damping is known from independent measurements. The comparison between precession and nutation spin waves is more intriguing in antiferromagnets: while 
the effective damping is identical in the two spin-wave bands in ferromagnets, in antiferromagnets 
nutation spin waves do not experience an exchange enhancement of the effective damping.

\section*{Acknowledgments}
The authors would like to thank Mikhail Cherkasskii for fruitful discussions.
We gratefully acknowledge the funding from the Swedish Research Council via Grant No. 2019-06313 and from the National Research, Development, and Innovation Office of Hungary under Project No. K131938.

\appendix

\begin{widetext}
\section{Derivation of the linearized ILLG equation for ferromagnets}
\label{AppendixA}

The Hamiltonian Eq.~\eqref{Eq4} in the harmonic approximation may be written as
\begin{align}
    \mathcal{H} & = E_{0}+H_{\textrm{lin}}+H_{\textrm{SW}},
    \label{Eq4b}
\end{align}
with 
\begin{align}
    E_{0} & =\frac{1}{2}\sum_{i\neq j}J_{ij}^{zz}+\sum_{i}K_{i}^{zz}-\sum_{i}B_{i}^{z}M_{i}\,
    \end{align}
    the ground-state energy,
    \begin{align}
H_{\textrm{lin}}=&\frac{1}{2}\sum_{i\neq j}\left(J_{ij}^{xz}\beta_{2i}+J_{ij}^{zx}\beta_{2j}-J_{ij}^{yz}\beta_{1i}-J_{ij}^{zy}\beta_{1j}\right)+\sum_{i}\left(2K_{i}^{xz}\beta_{2i}-2K_{i}^{yz}\beta_{1i}\right) -\sum_{i}M_{i}\left(B_{i}^{x}\beta_{2i}-B_{i}^{y}\beta_{1i}\right)\,
\end{align}
the term linear in the expansion variables $\beta_{1i}$ and $\beta_{2i}$, and
\begin{align}
    H_{\textrm{SW}}=&\frac{1}{2}\sum_{i\neq j}\left[-J_{ij}^{zz}\left(\dfrac{{\beta}_{1i}^2}{2} + \dfrac{{\beta}_{2i}^2}{2}+\dfrac{{\beta}_{1j}^2}{2} + \dfrac{{\beta}_{2j}^2}{2}\right)
    +J_{ij}^{xx}\beta_{2i}\beta_{2j}+J_{ij}^{yy}\beta_{1i}\beta_{1j}-J_{ij}^{xy}\beta_{2i}\beta_{1j}-J_{ij}^{yx}\beta_{1i}\beta_{2j}\right]\nonumber\\
    &+\sum_{i}\left[K_{i}^{xx}\beta_{2i}^2+K_{i}^{yy}\beta_{1i}^2-K_{i}^{zz}\left(\beta_{2i}^2+\beta_{1i}^2\right)\right] -\sum_i 2K_{i}^{xy}\beta_{1i}\beta_{2i}+\sum_{i}B_{i}^{z}\left(\dfrac{{\beta}_{1i}^2}{2} + \dfrac{{\beta}_{2i}^2}{2}\right)M_{i}\,.
\end{align}
the spin-wave Hamiltonian. The term $H_{\textrm{lin}}$ must vanish if the ferromagnetic state is equilibrium, leading to the conditions
\begin{align}
    \sum_{j}J_{ij}^{xz}+2K_{ij}^{xz}-B_{i}^{x}M_{i}=0,\label{Eqa}\\
    \sum_{j}J_{ij}^{yz}+2K_{ij}^{yz}-B_{i}^{y}M_{i}=0\label{Eqb},
\end{align}
when using the identity $J_{ij}^{\alpha\beta}=J_{ji}^{\beta\alpha}$ following from the definition of the Hamiltonian including a double summation over lattice sites $i$ and $j$.

The effective field in Eq.~\eqref{Eq2} is computed as
\begin{align}
    \bm{B}^{\rm eff}_i & = - \frac{1}{M_i} \frac{\partial \mathcal{H}}{\partial \boldsymbol{S}_i} = - \frac{1}{M_i} \frac{\partial \mathcal{H}}{\partial {\beta}_{2i}}\hat{\bm{x}} + \frac{1}{M_i} \frac{\partial \mathcal{H}}{\partial {\beta}_{1i}}\hat{\bm{y}}\,.
    \label{Eq6}
\end{align}
Using this expression, the linearized ILLG equation may be written as
\begin{align}
\label{EqA7}
    \frac{\mathrm{d} \beta_{2i}}{\mathrm{d}t}  =&  \frac{\gamma_{i}}{M_{i}}\sum_{j}\left(-J_{ij}^{zz}\beta_{1i}+J_{ij}^{yy}\beta_{1j}-J_{ij}^{yx}\beta_{2j}\right)\nonumber\\&+\frac{\gamma_{i}}{M_{i}}\left(-2K_{i}^{zz}\beta_{1i}+2K_{i}^{yy}\beta_{1i}-2K_{i}^{xy}\beta_{2i}\right) 
    +\gamma_{i}B_{i}^{z}\beta_{1i} + \alpha_i\frac{\mathrm{d}\beta_{1i}}{\mathrm{d}t} + \eta_i \frac{\mathrm{d}^2\beta_{1i}}{\mathrm{d}t^2},\\
    \frac{\mathrm{d} \beta_{1i}}{\mathrm{d}t}  =& -\frac{\gamma_{i}}{M_{i}}\sum_{j}\left(-J_{ij}^{zz}\beta_{2i}+J_{ij}^{xx}\beta_{2j}-J_{ij}^{xy}\beta_{1j}\right)\nonumber\\&-\frac{\gamma_{i}}{M_{i}}\left(-2K_{i}^{zz}\beta_{2i}+2K_{i}^{xx}\beta_{2i}-2K_{i}^{xy}\beta_{1i}\right)
     -\gamma_{i}B_{i}^{z} \beta_{2i} - \alpha_i \frac{\mathrm{d}\beta_{2i}}{\mathrm{d}t} - \eta_i \frac{\mathrm{d}^2\beta_{2i}}{\mathrm{d}t^2}.
    \label{EqA8}
\end{align}
For translationally invariant systems, Eqs.~\eqref{EqA7} and \eqref{EqA8} may be rewritten in a block-diagonal form in $\boldsymbol{k}$ after Fourier transformation,
\begin{align}
\label{EqA7a}
    \frac{\mathrm{d} \tilde{\beta}_{2}\left(\boldsymbol{k}\right)}{\mathrm{d}t}  =&  \frac{\gamma}{M}\left(-J_{\boldsymbol{k}=\boldsymbol{0}}^{zz}+J_{\boldsymbol{k}}^{yy}-2K^{zz}+2K^{yy}+MB^{z}\right)\tilde{\beta}_{1}\left(\boldsymbol{k}\right)+\frac{\gamma}{M}\left(-J_{\boldsymbol{k}}^{yx}-2K^{xy}\right)\tilde{\beta}_{2}\left(\boldsymbol{k}\right) 
     + \alpha\frac{\mathrm{d}\tilde{\beta}_{1}\left(\boldsymbol{k}\right)}{\mathrm{d}t} + \eta \frac{\mathrm{d}^2\tilde{\beta}_{1}\left(\boldsymbol{k}\right)}{\mathrm{d}t^2},\\
    \frac{\mathrm{d} \tilde{\beta}_{1}\left(\boldsymbol{k}\right)}{\mathrm{d}t}  =&-\frac{\gamma}{M}\left(-J_{\boldsymbol{k}}^{xy}-2K^{xy}\right)\tilde{\beta}_{1}\left(\boldsymbol{k}\right) -\frac{\gamma}{M}\left(-J_{\boldsymbol{k}=\boldsymbol{0}}^{zz}+J_{\boldsymbol{k}}^{xx}-2K^{zz}+2K^{xx}+MB^{z}\right)\tilde{\beta}_{2}\left(\boldsymbol{k}\right)
     - \alpha \frac{\mathrm{d}\tilde{\beta}_{2}\left(\boldsymbol{k}\right)}{\mathrm{d}t} - \eta \frac{\mathrm{d}^2\tilde{\beta}_{2}\left(\boldsymbol{k}\right)}{\mathrm{d}t^2},
    \label{EqA8a}
\end{align}
with the definitions
\begin{align}
    \tilde{\beta}_{1(2)}\left(\boldsymbol{k}\right) & = \frac{1}{\sqrt{N}} \sum_{\bm{R}_i} e^{-{\rm i} \bm{k}\cdot \bm{R}_i} \beta_{1i(2i)},\\
    \boldsymbol{J}_{\boldsymbol{k}} & =  \sum_{\bm{R}_i-\bm{R}_j} e^{-{\rm i} \bm{k}\cdot \left(\bm{R}_i-\bm{R}_j\right)} \boldsymbol{J}_{ij}.
    \label{Eq9}
\end{align}

Switching to the circularly polarized basis and performing Fourier transformation in time results in Eq.~\eqref{Eq10} in the main text, with the spin-wave Hamiltonian $H_{\textrm{SW}}\left(\bm{k}\right)$ given by
\begin{align}
    H_{\textrm{SW}}\left(\bm{k}\right)=&\begin{pmatrix}\Omega_{+}\left(\bm{k}\right) &  W_{+}\left(\bm{k}\right) \\  W_{-}\left(\bm{k}\right) & \Omega_{-}\left(\bm{k}\right)\end{pmatrix},\label{Eq10d}\\
    \Omega_{\pm}\left(\bm{k}\right)=&\frac{\gamma}{2M}\left(-2J_{\boldsymbol{k}=\boldsymbol{0}}^{zz}+J_{\boldsymbol{k}}^{xx}+J_{\boldsymbol{k}}^{yy}\mp\textrm{i}J_{\boldsymbol{k}}^{yx}\pm\textrm{i}J_{\boldsymbol{k}}^{xy}-4K^{zz}+2K^{xx}+2K^{yy}+2MB^{z}\right),\label{Eq18}\\
    W_{\pm}\left(\bm{k}\right)=&\frac{\gamma}{2M}\left(J_{\boldsymbol{k}}^{xx}-J_{\boldsymbol{k}}^{yy}\mp\textrm{i}J_{\boldsymbol{k}}^{yx}\mp\textrm{i}J_{\boldsymbol{k}}^{xy}+2K^{xx}-2K^{yy}\mp\textrm{i}4K^{xy}\right).\label{Eq19}
\end{align}
The relations $\Omega_{\pm}^{*}\left(\boldsymbol{k}\right)=\Omega_{\mp}\left(-\boldsymbol{k}\right)$ and $W_{\pm}^{*}\left(\boldsymbol{k}\right)=W_{\mp}\left(-\boldsymbol{k}\right)$ are satisfied by definition, and they enforce the particle-hole symmetry mentioned in the main text.

To calculate the dependence of the gyromagnetic ratio on the wave vector in biaxial ferromagnets, we keep the off-diagonal terms $W_{\pm}\left(\bm{k}\right)$ of $H_{\textrm{SW}}$ in Eq.~\eqref{Eq18}. In the absence of an inertial term, this yields the dispersion relation
\begin{align}
    &\omega^{(0)}_{\rm Prec}(\bm{k})=\frac{1}{1+\alpha^2}\left(\frac{\Omega_{-}-\Omega_{+}+\textrm{i}\alpha\left(\Omega_{-}+\Omega_{+}\right)}{2}+\sqrt{\left(1-\alpha^{2}\right)\left(\frac{\Omega_{-}+\Omega_{+}}{2}\right)^{2}+\textrm{i}\alpha\frac{\Omega^{2}_{-}-\Omega^{2}_{+}}{2}-W_{+}W_{-}}\right),
\end{align}
the gyromagnetic ratio
\begin{align}
\gamma^{(0)}_{\rm eff}(\bm{k})=\frac{\gamma}{\sqrt{1-\frac{4W_{+}(\bm{k})W_{-}(\bm{k})}{\left(\Omega_{-}(\bm{k})+\Omega_{+}(\bm{k})\right)^{2}}}}\label{Eff_Gyro0}
\end{align}
for $\alpha=0$, and the effective damping parameter
\begin{align}
\alpha^{(0)}_{\rm eff}(\bm{k})=\frac{\alpha}{\sqrt{1-\frac{4W_{+}(\bm{k})W_{-}(\bm{k})}{\left(\Omega_{-}(\bm{k})+\Omega_{+}(\bm{k})\right)^{2}}}}.\label{Eff_alpha0}
\end{align}
While Eq.~\eqref{Eff_Gyro0} predicts a decrease of the gyromagnetic ratio with $\boldsymbol{k}$ similarly to Eq.~\eqref{Eff_Gyro}, quantitatively the formulae depend differently on the wave vector, as discussed in the main text. The same conclusion applies to the effective damping, since Eq.~\eqref{Eff_alpha0} shows the same dependence on the wave vector as Eq.~\eqref{Eff_Gyro0}, similarly to Eqs.~\eqref{alphaeff} and \eqref{Eff_Gyro} in the inertial case.

\section{Derivation of the linearized ILLG equation for antiferromagnets}
\label{AppendixB}
Using the definition of the spins on the two sublattices in Eqs. (\ref{Eq_anti1}) and (\ref{Eq_anti2}), expansion of the Hamiltonian for an antiferromagnet has the following form: 
\begin{align}
    E_{0}=&\sum_{i\in A, j\in B}-J_{ij}^{zz}+\sum_{i\in A}K_{i}^{zz}+\sum_{j\in B}K_{j}^{zz}-\sum_{i\in A}B_{i}^{z}M_{i,A}-\sum_{j\in B}-B_{j}^{z}M_{j,B}\,,\\
H_{\textrm{lin}}=&\sum_{i\in A, j\in B}\left(-J_{ij}^{xz}\beta_{2i}+J_{ij}^{zx}\beta_{2j}+J_{ij}^{yz}\beta_{1i}+J_{ij}^{zy}\beta_{1j}\right)+\sum_{i\in A}\left(2K_{i}^{xz}\beta_{2i}-2K_{i}^{yz}\beta_{1i}\right)+\sum_{j\in B}\left(-2K_{j}^{xz}\beta_{2j}-2K_{j}^{yz}\beta_{1j}\right)\nonumber\\&-\sum_{i\in A}M_{i,A}\left(B_{i}^{x}\beta_{2i}-B_{i}^{y}\beta_{1i}\right)-\sum_{j\in B}M_{j,B}\left(B_{i}^{x}\beta_{2i}+B_{i}^{y}\beta_{1i}\right)\,,\\
    H_{\textrm{SW}} =&\sum_{i\in A, j\in B}\left[J_{ij}^{zz}\left(\dfrac{{\beta}_{1i}^2}{2} + \dfrac{{\beta}_{2i}^2}{2}+\dfrac{{\beta}_{1j}^2}{2} + \dfrac{{\beta}_{2j}^2}{2}\right)+J_{ij}^{xx}\beta_{2i}\beta_{2j}-J_{ij}^{yy}\beta_{1i}\beta_{1j}+J_{ij}^{xy}\beta_{2i}\beta_{1j}-J_{ij}^{yx}\beta_{1i}\beta_{2j}\right]\nonumber\\&+\sum_{i\in A}\left(K_{i}^{xx}\beta_{2i}^2+K_{i}^{yy}\beta_{1i}^2-K_{i}^{zz}\left(\beta_{2i}^2+\beta_{1i}^2\right)-2K_{i}^{xy}\beta_{1i}\beta_{2i}\right)\nonumber\\&+\sum_{j\in B}\left(K_{j}^{xx}\beta_{2j}^2+K_{j}^{yy}\beta_{1j}^2-K_{j}^{zz}\left(\beta_{2j}^2+\beta_{1j}^2\right)+2K_{j}^{xy}\beta_{1j}\beta_{2j}\right)\nonumber\\&+\sum_{i\in A}B_{i}^{z}\left(\dfrac{{\beta}_{1i}^2}{2} + \dfrac{{\beta}_{2i}^2}{2}\right)M_{i,A}+\sum_{j\in B}-B_{j}^{z}\left(\dfrac{{\beta}_{1j}^2}{2} + \dfrac{{\beta}_{2j}^2}{2}\right)M_{j,B}\label{Hamil_Antiferro}
\end{align}

The linearized equations of motion are calculated similarly to the ferromagnetic case. The effective fields acting on each sublattice $\bm{B}^{\rm eff,A}_i$ and $\bm{B}^{\rm eff,B}_j$ are determined and inserted into the ILLG equation. Translational invariance is assumed with $\gamma_{i,A/B}=\gamma_{A/B}$, $\alpha_{i,A/B}=\alpha_{A/B}$, $\eta_{i,A/B}=\eta_{A/B}$, $M_{i,A/B}=M_{A/B}$, $\boldsymbol{K}_{i/j}=\boldsymbol{K}_{A/B}$ and $\boldsymbol{J}_{ij}=\boldsymbol{J}\left(\boldsymbol{R}_{i}-\boldsymbol{R}_{j}\right)$, enabling diagonalization in lattice sites via Fourier transformation. In the circularly polarized basis and after Fourier transformation in time, the equations of motion are formally identical to Eq.~\eqref{Eq10}, as mentioned in the main text. The spin-wave Hamiltonian reads
\begin{align}
    H_{\textrm{SW}}\left(\bm{k}\right)=&
\begin{pmatrix}
    \Omega_{+AA}\left(\boldsymbol{k}\right) & W_{+AA}\left(\bm{k}\right) & \Omega_{+AB}\left(\boldsymbol{k}\right) & W_{+AB}\left(\bm{k}\right) \\
    W_{-AA}\left(\bm{k}\right) & \Omega_{-AA}\left(\boldsymbol{k}\right)  & W_{-AB}\left(\bm{k}\right) & \Omega_{-AB}\left(\boldsymbol{k}\right) \\
     \Omega_{+BA}\left(\boldsymbol{k}\right) & W_{+BA}\left(\bm{k}\right) & \Omega_{+BB}\left(\boldsymbol{k}\right) & W_{+BB}\left(\bm{k}\right) \\
     W_{-BA}\left(\bm{k}\right) & \Omega_{-BA}\left(\boldsymbol{k}\right) &    W_{-BB}\left(\bm{k}\right) & \Omega_{-BB}\left(\boldsymbol{k}\right) 
    \end{pmatrix},
    \label{EqAFM}
\end{align}
containing the coefficients
\begin{align}
     \Omega_{\pm AA}\left(\bm{k}\right)= &\frac{\gamma_{A}}{2M_{A}}\left(J_{\boldsymbol{k}AA}^{xx}+J_{\boldsymbol{k}AA}^{yy}\mp\textrm{i}J_{\boldsymbol{k}AA}^{yx}\pm\textrm{i}J_{\boldsymbol{k}AA}^{xy}-2J_{\boldsymbol{k}=\boldsymbol{0}AA}^{zz}+2J_{\boldsymbol{k}=\boldsymbol{0}AB}^{zz}-4K_{A}^{zz}+2K_{A}^{xx}+2K_{A}^{yy}+2M_{A}B^{z}\right),\\
    W_{\pm AA}\left(\bm{k}\right)= &\frac{\gamma_{A}}{2M_{A}}\left(J_{\boldsymbol{k}AA}^{xx}-J_{\boldsymbol{k}AA}^{yy}\mp\textrm{i}J_{\boldsymbol{k}AA}^{yx}\mp\textrm{i}J_{\boldsymbol{k}AA}^{xy}+2K_{A}^{xx}-2K_{A}^{yy} \mp \textrm{i}4K_{A}^{xy}\right),\\
    \Omega_{\pm BB}\left(\bm{k}\right)= &\frac{\gamma_{B}}{2M_{B}}\left(J_{\boldsymbol{k}BB}^{xx}+J_{\boldsymbol{k}BB}^{yy}\pm\textrm{i}J_{\boldsymbol{k}BB}^{yx}\mp\textrm{i}J_{\boldsymbol{k}BB}^{xy}-2J_{\boldsymbol{k}=\boldsymbol{0}BB}^{zz}+2J_{\boldsymbol{k}=\boldsymbol{0}BA}^{zz}-4K_{B}^{zz}+2K_{B}^{xx}+2K_{B}^{yy}-2M_{B}B^{z}\right),\\
    W_{\pm BB}\left(\bm{k}\right)= &\frac{\gamma_{B}}{2M_{B}}\left(J_{\boldsymbol{k}BB}^{xx}-J_{\boldsymbol{k}BB}^{yy}\pm\textrm{i}J_{\boldsymbol{k}BB}^{yx}\pm\textrm{i}J_{\boldsymbol{k}BB}^{xy}+2K_{B}^{xx}-2K_{B}^{yy} \pm \textrm{i}4K_{B}^{xy}\right),\\
    \Omega_{\pm AB}\left(\bm{k}\right)= &\frac{\gamma_{A}}{2M_{A}}\left(-J_{\boldsymbol{k}AB}^{yy}+J_{\boldsymbol{k}AB}^{xx}\mp \textrm{i}J_{\boldsymbol{k}AB}^{yx}\mp \textrm{i}J_{\boldsymbol{k}AB}^{xy}\right),\\
    W_{\pm AB}\left(\bm{k}\right)= &\frac{\gamma_{A}}{2M_{A}}\left(J_{\boldsymbol{k}AB}^{yy}+J_{\boldsymbol{k}AB}^{xx}\mp \textrm{i}J_{\boldsymbol{k}AB}^{yx}\pm \textrm{i}J_{\boldsymbol{k}AB}^{xy}\right),\\
    \Omega_{\pm BA}\left(\bm{k}\right)= &\frac{\gamma_{B}}{2M_{B}}\left(-J_{\boldsymbol{k}BA}^{yy}+J_{\boldsymbol{k}BA}^{xx}\pm \textrm{i}J_{\boldsymbol{k}BA}^{yx}\pm \textrm{i}J_{\boldsymbol{k}BA}^{xy}\right),\\
    W_{\pm BA}\left(\bm{k}\right)= &\frac{\gamma_{B}}{2M_{B}}\left(J_{\boldsymbol{k}BA}^{yy}+J_{\boldsymbol{k}BA}^{xx}\pm \textrm{i}J_{\boldsymbol{k}BA}^{yx}\mp \textrm{i}J_{\boldsymbol{k}BA}^{xy}\right).
\end{align}

Particle-hole symmetry is enforced by the coefficients satisfying $\Omega_{\pm}^{*}\left(\boldsymbol{k}\right)=\Omega_{\mp}\left(-\boldsymbol{k}\right)$ and $W_{\pm}^{*}\left(\boldsymbol{k}\right)=W_{\mp}\left(-\boldsymbol{k}\right)$ for all sublattice indices. 

Assuming that the sublattices are identical and the external field is set to zero, the spin-wave Hamiltonian simplifies to
\begin{align}
    H_{\textrm{SW}}\left(\bm{k}\right)=&
\begin{pmatrix}
    \Omega\left(\boldsymbol{k}\right) & W\left(\bm{k}\right) & \Omega^{'}\left(\boldsymbol{k}\right) & W^{'}\left(\bm{k}\right) \\
    W^{*}\left(-\bm{k}\right) & \Omega\left(-\boldsymbol{k}\right)  & W^{'*}\left(-\bm{k}\right) & \Omega^{'*}\left(-\boldsymbol{k}\right) \\
     \Omega^{'*}\left(\boldsymbol{k}\right) & W^{'}\left(-\bm{k}\right) & \Omega\left(-\boldsymbol{k}\right) & W^{*}\left(-\bm{k}\right) \\
     W^{'*}\left(\bm{k}\right) & \Omega^{'}\left(-\boldsymbol{k}\right) &    W\left(\bm{k}\right) & \Omega\left(\boldsymbol{k}\right) 
    \end{pmatrix},
\end{align}
where $\Omega\left(\boldsymbol{k}\right)=\Omega^{*}\left(\boldsymbol{k}\right)$ and $W\left(\bm{k}\right)=W\left(-\bm{k}\right)$. Time-reversal symmetry with $\mathcal{T}=\tau^{x}\mathcal{K}$ also requires $W^{'}\left(\bm{k}\right)=W^{'*}\left(\bm{k}\right)$ and $\Omega^{'}\left(\bm{k}\right)=\Omega^{'}\left(-\bm{k}\right)$. These conditions are satisfied if the two sublattices together form a Bravais lattice, e.g. the centered rectangular lattice consisting of two primitive rectangular sublattices on the bcc(110) surface discussed in the main text. Such effective time-reversal symmetries in antiferromagnetic systems are always protected by a crystal symmetry, since the physical time reversal inverts the spin direction on both sublattices, and the crystal symmetry is required for exchanging the two sublattices in order to get back to the initial configuration.

For the monolayer on the bcc(110) surface discussed in the main text, any two lattice sites can be exchanged by a $180^{\circ}$ rotation around the out-of-plane $y$ axis, which implies $J_{ij}^{xy}=-J_{ij}^{yx}$. The mirror symmetry on the $yz$ plane leads to $K^{xy}=0$. Due to these conditions, $H_{\textrm{SW}}\left(\bm{k}\right)$ is a real matrix (in particular, $\Omega^{'}\left(\boldsymbol{k}\right)=\Omega^{'*}\left(\boldsymbol{k}\right)$ and $W\left(\bm{k}\right)=W^{*}\left(\bm{k}\right)$ since the other two coefficients are real anyways as mentioned above), and the equation of motion satisfies reciprocal symmetry with $\mathcal{R}=\tau^{x}$, as discussed in the main text.

 To calculate the effective gyromagnetic ratios and damping parameters, we treat the damping $\alpha$ and the magnetic field $B^{z}$ perturbatively in Eq.~\eqref{Secular_AFM}. For identical sublattices this yields
\begin{align}
&\eta^{2}\omega^{4}\left(\boldsymbol{k}\right)-\left(1+2b\right)\omega^{2}\left(\boldsymbol{k}\right)+\Omega^{2}-W_{\pm}^{2}\left(\boldsymbol{k}\right)+\textrm{i}2\alpha \left(\Omega\omega'\left(\boldsymbol{k}\right)-\eta\left(\omega'\left(\boldsymbol{k}\right)\right)^{3}\right)\mp 2\gamma\omega'\left(\boldsymbol{k}\right)B^{z}=0,
\end{align}
where $\omega'\left(\boldsymbol{k}\right)$ is the eigenfrequency without damping or magnetic field. Following the procedure in Ref.~\cite{Mondal2020nutation}, the eigenfrequencies are extended up to linear order in $B^{z}$ and $\alpha$, leading to Eqs.~\eqref{gamma_AFM}-\eqref{alpha_AFM_Nut} in the main text.

\end{widetext}


%

\end{document}